\documentclass[aps, reprint, prl, twocolumn, superscriptaddress,amsmath,amssymb,noeprint]{revtex4-2}

\usepackage{times}
\usepackage{amsmath}
\usepackage{amssymb}
\usepackage{xfrac}
\usepackage{dsfont}
\usepackage{graphicx}
\usepackage{braket}
\usepackage{bbold}
\usepackage[normalem]{ulem}
\usepackage{bm}
\usepackage{bbm}
\usepackage{color}
\usepackage{xcolor}
\usepackage{pdfpages}
\usepackage{pgffor}
\usepackage{float}

\makeatletter
\AtBeginDocument{\let\LS@rot\@undefined}
\makeatother

\usepackage[colorlinks=true, urlcolor=blue, linkcolor=blue, citecolor=blue]{hyperref}
\usepackage{siunitx}

\newcommand{\berkeleyphy}{Department of Physics, University of California, Berkeley, California 94720}
\newcommand{\CIQC}{Challenge Institute for Quantum Computation, University of California, Berkeley, California 94720}
\newcommand{\LBL}{Materials Sciences Division, Lawrence Berkeley National Laboratory, Berkeley, California 94720}

\begin{document}


\title{Higher symmetry breaking and non-reciprocity in a driven-dissipative Dicke model}

\author{Jacquelyn Ho}
\affiliation{\berkeleyphy}
\affiliation{\CIQC}

\author{Yue-Hui Lu}
\affiliation{\berkeleyphy}
\affiliation{\CIQC}

\author{Tai Xiang}
\affiliation{\berkeleyphy}
\affiliation{\CIQC}

\author{Tsai-Chen Lee}
\affiliation{\berkeleyphy}
\affiliation{\CIQC}

\author{Zhenjie Yan}
\affiliation{\berkeleyphy}
\affiliation{\CIQC}

\author{Dan M. Stamper-Kurn}
\email[]{dmsk@berkeley.edu}
\affiliation{\berkeleyphy}
\affiliation{\CIQC}
\affiliation{\LBL}
\begin{abstract}

Higher symmetries in interacting many-body systems often give rise to new phases and unexpected dynamical behavior. Here, we theoretically investigate a variant of the Dicke model with higher-order discrete symmetry, resulting from complex-valued coupling coefficients between quantum emitters and a bosonic mode. We propose a driven-dissipative realization of this model focusing on optomechanical response of a driven atom tweezer array comprised of $n$ sub-ensembles and placed within an optical cavity, with the phase of the driving field advancing stepwise between sub-ensembles.  Examining stationary points and their dynamical stability, we identify a phase diagram for $n\geq 3$ with three distinctive features: a $\mathbb{Z}_n$ ($\mathbb{Z}_{2n}$) symmetry-breaking superradiant phase for even (odd) $n$, a normal unbroken-symmetry phase that is dynamically unstable due to non-reciprocal forces between emitters, and a first-order phase transition separating these phases. This $n$-phase Dicke model may be equivalently realized in a variety of optomechanical or opto-magnonic settings, where it can serve as a testbed for studying high-order symmetry breaking and non-reciprocal interactions in open systems.

\end{abstract}

\maketitle

Phase transitions of interacting many-body systems at thermal equilibrium are a long-standing topic of research.  More recently, phase transitions in driven-dissipative many-body quantum systems have been explored~\cite{hinrichsenNonequilibriumPhaseTransitions2006,mathenyExoticStatesSimple2019,baumannExploringSymmetryBreaking2011,youngNonequilibriumFixedPoints2020,szymanskaNonequilibriumQuantumCondensation2006,belyanskyPhaseTransitionsNonreciprocal2025}, opening the question of whether emergent properties of such open systems differ from those found at equilibrium.

The Dicke model \cite{dickeCoherenceSpontaneousRadiation1954, heppSuperradiantPhaseTransition1973} is paradigmatic for the study of both equilibrium and driven-dissipative phase transitions.  It has played a defining role in studies of quantum optical phenomena such as superradiance and lasing~\cite{dickeCoherenceSpontaneousRadiation1954,schaferContinuousRecoildrivenLasing2025,norciaSuperradianceMillihertzLinewidth2016,meiserProspectsMillihertzLinewidthLaser2009} as well as in quantum simulation of phase transitions in macroscopic~\cite{baumannDickeQuantumPhase2010,mivehvarCavityQEDQuantum2021,ritschColdAtomsCavitygenerated2013,safavi-nainiVerificationManyIonSimulator2018} and mesoscopic~\cite{hoOptomechanicalSelforganizationMesoscopic2025a} systems.  The model describes a collection of $N$ two-level quantum emitters identically coupled to a single-mode electromagnetic field.  As a closed system, the Dicke model undergoes a second-order phase transition as the emitter-mode coupling strength increases, going from an unbroken-symmetry normal phase to a $\mathbb{Z}_2$ symmetry-breaking superradiant phase.

Driven-dissipative versions of the Dicke model have been realized experimentally.  Prominent among these are systems involving atomic ensembles placed within a high-finesse optical cavity \cite{baumannDickeQuantumPhase2010, zhiqiangNonequilibriumPhaseTransition2017}. The two-level quantum emitter of the original Dicke model is replaced by optically driven atoms undergoing low-energy Bragg or Raman transitions, between different mechanical or hyperfine-spin states, respectively.  Under certain conditions, the driven-dissipative system reaches a steady state.  As in the equilibrium Dicke model, the steady-state shows a normal-to-superradiant phase transition.

Since the Dicke model considers spins that are all symmetrically coupled to the cavity, one might ask how the phenomenology changes when this condition is broken. This modification of the Dicke model has been studied in the context of disorder, where different spins have different energy-level spacings and can couple to the cavity with different strengths~\cite{nairnSpinSelfOrganizationOptical2025,temnovSuperradianceSubradianceInhomogeneously2005,dasDickeModelDisordered2024,zhangDickemodelSimulationCavityassisted2018} and with a continuum of phases~\cite{zhangDickemodelSimulationCavityassisted2018}. Disordered couplings among atoms placed in a multimode cavity have also been used to realize glassy systems~\cite{kroezeDirectlyObservingReplica2025,marshEntanglementReplicaSymmetry2024}. In contrast, there has been little exploration of what instead happens when the complex spin-cavity coupling varies \textit{discretely} from spin to spin as $\sum_{i=1}^N (\chi_i^*\hat{c}+\hat{c}^\dagger \chi_i)\hat{\sigma}_i^x$, where $\hat{c}$ is the electromagnetic mode field operator, $\bm{\sigma}_i$ is the pseudo-spin 1/2 vector for emitter $i$, and $\chi_i$ is a complex-valued emitter-mode coupling strength.  We find that in this scenario, it is possible to generate systems with discrete symmetries higher than $\mathbb{Z}_2$ with a simple modification. Specifically, we consider that $N$ emitters are divided into $n$ groups indexed by $j$, and the phase of the spin-cavity coupling for each group is set to $2\pi j/n$. Similar forms of this discrete symmetry have been previously explored in Ising-like systems such as the Potts model~\cite{wuPottsModel1982} and the \textit{q}-state clock model~\cite{chatterjeeOrderingKineticsState2018}.  Generalizing from the $\mathbb{Z}_2$ parity symmetry of the canonical Dicke model, the system we consider now has $\mathbb{Z}_n$ symmetry for even $n$ and $\mathbb{Z}_{2n}$ symmetry for odd $n$. We call this the $n$-phase Dicke model.

\begin{figure*}
    \centering
    \includegraphics{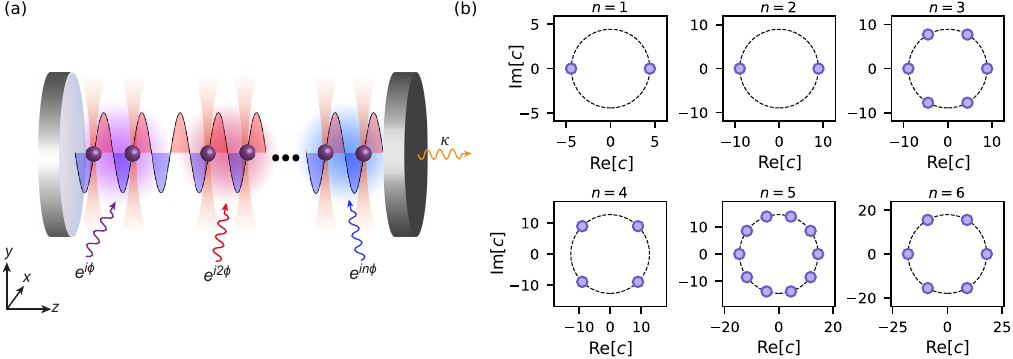}
    \caption{(a) Schematic of the setup for the $n$-phase Dicke model. Atoms are held in harmonic traps at the nodes of the cavity field. The atoms are divided into $n$ groups and each group is illuminated by a pump laser with a $\phi=2\pi/n$ phase difference between adjacent groups. (b) Real and imaginary quadratures of the steady-state cavity field (solid circles) calculated using the atomic center-of-mass positions found through Lyapunov function minimization, shown for $n=1$ through $n=6$. The results show $\mathbb{Z}_n$ symmetry for even $n$ and $\mathbb{Z}_{2n}$ symmetry for odd $n$. The parameters used are $\nu=30$, $\omega_z=2\pi\times 70$~kHz, $\mathrm{\Delta_{pa}}=-2\pi\times 100$~MHz, $\mathrm{\Delta_{pc}}=-2\pi\times 4$~MHz, $\kappa=0$, $g_0=2\pi\times 3$~MHz, and $\mathrm{\Omega}=2\pi\times20$~MHz. }
    \label{fig:M1}
\end{figure*}

We propose a way to realize the $n$-phase Dicke model using ultracold atoms trapped in optical tweezers, though we expect the concepts to be applicable to other realizations. We specifically consider a pumped and dissipative form of the Dicke model that relies on optomechanical self-organization of the atoms~\cite{ritschColdAtomsCavitygenerated2013,mivehvarCavityQEDQuantum2021}. Here, the superradiant phase corresponds to the atoms self-organizing onto the lattice formed by the interference of the pump and cavity fields, and the atomic motion can be mapped to spins. Realization of the aforementioned $\mathbb{Z}_n$ or $\mathbb{Z}_{2n}$ symmetry occurs by adjusting separately the optical phase of the pump that drives each atom.  

Analyzing this system theoretically, we characterize a phase diagram with three novel features.  First, we find conditions at high pump strength where the cavity field and emitters together stably break the discrete $\mathbb{Z}_n$ or $\mathbb{Z}_{2n}$ symmetry, for even and odd $n$, respectively.  Second, unlike in the canonical Dicke model, the phase transition to the stable broken symmetry state becomes first-order, asymptoting to a second-order transition only when the pump light is far-detuned from the cavity resonance frequency.  Third, we find the normal phase to be dynamically unstable. This instability arises from non-reciprocal light-mediated interactions between the driven emitters.

\textit{Cavity optomechanical system---}We  elucidate the features of the $n$-phase Dicke model in the specific experimental context of cavity optomechanics with an atom tweezer array \cite{hoOptomechanicalSelforganizationMesoscopic2025a,wangProgrammableFewatomBragg2025}.  We consider the one-dimensional motion of a one-dimensional array of $N$ harmonically confined $^{87}$Rb atoms along the axis ($\mathbf{z}$) of a single-mode Fabry-P\'erot optical cavity. The harmonic traps are centered on the cavity nodes and spaced by an integer number of wavelengths. The atoms, which have an allowed dipole transition at optical frequency  $\omega_\mathrm{a}$, are then pumped  with coherent light propagating transverse to the cavity axis (along $\mathbf{x}$) at frequency $\omega_\mathrm{p}$ (wavenumber $k$ and wavelength $\lambda$) and Rabi frequency $\mathrm{\Omega}$.   

When $\mathrm{\Omega}$ is uniform across all atoms, an optomechanical phase transition occurs at a critical pump strength $\mathrm{\Omega_c}$, related to the canonical Dicke phase transition, owing to competition between the trapping potentials, which confine atoms to the cavity-field nodes, and the interference of the pump and cavity fields, which pulls atoms away from the nodes. A choice of gauge allows us to define $\mathrm{\Omega}$ as being real and non-negative.  With $\mathrm{\Omega}<\mathrm{\Omega_c}$, the atoms remain stably centered on the cavity nodes and do not scatter light coherently into the cavity; this corresponds to the normal phase of the Dicke model.  With $\mathrm{\Omega}>\mathrm{\Omega_c}$, the atoms self-organize by moving collectively towards cavity antinodes with the same phase, where they emit coherently into the cavity.  This superradiant mode breaks the system's $\mathbb{Z}_2$ symmetry, with the atoms displaced either toward the positive- or negative-signed cavity field antinodes, and the cavity light corresondingly having a phase of either $0$ or $\pi$ with respect to the pump field \cite{hoOptomechanicalSelforganizationMesoscopic2025a}.

To generate higher-order symmetry, we now modify this setup  as shown in Fig.~\ref{fig:M1}(a). We suppose that a total of $N = n \nu$ trapped atoms, still each positioned on a cavity-field node, are divided into $n$ groups.  Each group $j$, containing $\nu$ atoms, is driven with a pump Rabi frequency of $\mathrm{\Omega}e^{i\phi j}$, where $\phi=2\pi/n$ and $j=\{1,2,...,n\}$.  We assume the pump is far detuned from the atomic resonance, i.e.\ that the absolute value $|\mathrm{\Delta_{pa}}| \equiv |\omega_\mathrm{p}-\omega_\mathrm{a}|$ is much greater than both $\mathrm{\Omega}$ and the excited-state decay rate, allowing us to adiabatically eliminate the atomic excited state and focus on dispersive atom-light interactions; this makes our treatment equivalent to that of other optomechanical systems of polarizable media. The Hamiltonian in the frame rotating at $\omega_\mathrm{p}$ reads as
\begin{flalign}
    \label{H}
    \hat{H}=-\hbar\mathrm{\Delta_{pc}}\hat{c}^\dagger\hat{c}&+\sum_{j=1}^n \Big[ \frac{\hat{p}_j^2}{2\mu}+\frac{\mu\omega_z^2\hat{z}_j^2}{2}\\\nonumber
    +&\frac{\hbar\mathrm{\Omega}\nu}{\mathrm{\Delta_{pa}}}g_0\sin(k\hat{z}_j)(e^{-i \phi j}\hat{c}+e^{i \phi j}\hat{c}^\dagger)\Big],
\end{flalign}
where $\omega_z$ is the tweezer trap frequency, $g_0$ is the vacuum Rabi frequency, and we have defined $\mu \equiv \nu m$ ($m$ being the atomic mass) and $\mathrm{\Delta_{pc}}\equiv\omega_\mathrm{p}-\omega_\mathrm{c}$. For simplicity, we include only the center-of-mass mode $\hat{z}_j$ of each group and ignore other modes of motion. We also ignore the dispersive shift of the cavity resonance by the atoms. Note that for $n=1$, Eq.~(\ref{H}) can be mapped to the canonical Dicke model in the Holstein-Primakoff representation~\cite{holsteinFieldDependenceIntrinsic1940} by re-writing the position and momentum operators in terms of bosonic creation and annihilation operators for the center-of-mass mode of the array.

We highlight the symmetries of the system: $\hat{H}$ preserves $\mathbb{Z}_n$ symmetry, since performing the gauge transformation $c\rightarrow e^{i\phi}c$ and simultaneous permutation of the group index $j$ leaves $\hat{H}$ invariant. The system also has a $\mathbb{Z}_2$ symmetry in that the simultaneous transformations $\{z,p\}\rightarrow \{-z,-p\}$ and $c\rightarrow-c$ also lead to the same Hamiltonian. These combined symmetries result in $\mathbb{Z}_{2n}$ symmetry for odd $n$ and $\mathbb{Z}_n$ symmetry for even $n$. In the even $n$ case, the $\mathbb{Z}_2$ symmetry is redundant because it is equivalent to permuting the group index $n/2$ times while advancing the cavity field phase by $n\phi/2=\pi$.  

In the following, we neglect stochastic noise on the cavity field and study the evolution of the expectation values of operators under the master equation. We obtain the following equations of motion:
\begin{align}
    \label{chatdot}
    \dot{c} &= (i\mathrm{\Delta_{pc}}-\kappa)c-i\sum_{l=1}^n\frac{\nu\mathrm{\Omega}}{\mathrm{\Delta_{pa}}}e^{i\phi l}g_0\sin(kz_l), \\
    \label{phatdot}
    \dot{p}_j &= -\mu\omega_z^2z_j-\frac{\nu\mathrm{\Omega}}{\mathrm{\Delta_{pa}}}\hbar kg_0\cos(kz_j)\big(e^{i\phi j}c^*+e^{-i\phi j}c\big)
    , \\
    \label{zhatdot}
    \dot{z}_j &= \frac{p_j}{\mu}.
\end{align}
Here, $\kappa$ is the decay rate of the cavity field. We can further reduce the equations of motion to a set of $2n$ purely mechanical equations by assuming that the cavity field is in its steady state and instantaneously follows the atomic motion. The adiabatic cavity field is described by 
\begin{eqnarray}
    \label{c}
    c=\frac{i\nu\mathrm{\Omega}}{\mathrm{\Delta_{pa}}}\frac{1}{i\mathrm{\Delta_{pc}}-\kappa}\sum_{j=1}^n e^{i\phi j}g_0\sin(kz_j).
\end{eqnarray}
From Eq.~(\ref{c}), one can also understand the symmetry-broken states as the sets of positions $\{z_j\simeq\pm\lambda/4\}$ that maximize the magnitude of $\sum_{j=1}^n {e^{i\phi j}\sin(kz_j)}$. For even $n$, there are $n$ such sets and for odd $n$ there are $2n$. We verify these $\mathbb{Z}_{n}$ and $\mathbb{Z}_{2n}$ symmetries by performing Lyapunov function minimization in the limit $\kappa=0$ [Fig.~\ref{fig:M1}(b)] to find all the symmetry-broken steady states~\cite{methods,schaefferOrdinaryDifferentialEquations2016,endresSimplicialHomologyAlgorithm2018}. Substituting Eq.~(\ref{c}) into Eq.~(\ref{phatdot}) yields the force equations
\begin{flalign}
    \label{EOMs}
    \dot{p}_j=-\mu\omega_z^2z_j-\frac{2\hbar\nu^2\mathrm{\Omega}^2}{\mathrm{\Delta_{pa}}^2}&\frac{g_0^2k\cos(kz_j)}{{\mathrm{\Delta_{pc}}^2+\kappa^2}}\sum_{l=1}^n\Big[\mathrm{\Delta_{pc}}\cos(\phi(j-l))&&\\\nonumber
    &-\kappa\sin(\phi(j-l))\Big]\sin(kz_l).&&
\end{flalign}

\begin{figure}
    \centering
    \includegraphics{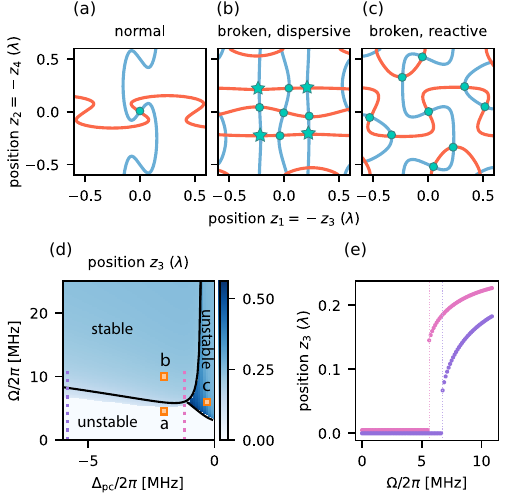}
    \caption{(a-c) Contour plots of the forces $\dot{p}_1=0$ (blue) and $\dot{p}_2=0$ (red) for $n=4$, where we have set $z_1=-z_3$ and $z_2=-z_4$. Steady states (teal markers) occur where the blue and red curves intersect. Jacobian eigenvalue analysis is used to determine whether each steady state is unstable (circles) or stable (stars). (d) Phase diagram for $n=4$ showing the steady-state position of one group ($z_3$) at a particular symmetry-broken solution. Orange squares correspond to the parameters used in (a-c). Black lines are phase boundaries calculated by expanding Eq.~(\ref{EOMs}) to third order around the symmetry-broken solutions. (e) Line cuts corresponding to the dashed lines in (d) showing the first-order transition as a function of $\mathrm{\Omega}$. Dashed lines show the location of the discontinuity for the two values of $\mathrm{\Delta_{pc}}$. Pink points have a slight vertical offset for visibility. In the $|\mathrm{\Delta_{pc}}|\rightarrow \infty$ limit, the transition becomes second-order (continuous). Calculations in figure are done with $\mathrm{\Delta_{pa}}=-2\pi\times 100$~MHz, $\kappa=2\pi\times 0.5$~MHz, $\nu=30$, $g_0=2\pi\times3$~MHz, and $\omega_z=2\pi\times 70$~kHz.}
    \label{fig:M2}
\end{figure}
\textit{Phase diagram of the $n$-phase Dicke model---}
To elucidate the phase diagram of the $n$-phase Dicke model, we first consider the infinitely dispersive cavity limit ($\mathrm{\Delta_{pc}}/\kappa\rightarrow-\infty$), in which the interactions are dominated by the ``cavity-dispersive" terms proportional to $\mathrm{\Delta_{pc}}\cos(\phi(j-l))$. In this limit, the ``cavity-reactive" terms proportional to $\kappa\sin(\phi(j-l))$ in Eq.~(\ref{EOMs}) are negligible and can be ignored. 
The phase diagram is now analogous to that of the Dicke model: There is a stable normal phase with unbroken symmetry corresponding to $z_j=0$ for all $j$ when $\mathrm{\Omega}<\mathrm{\Omega_c}$, which continuously transitions into a superradiant, broken-symmetry phase at $\mathrm{\Omega}=\mathrm{\Omega_c}$, where $\mathrm{\Omega_c}=\sqrt{\frac{\mathrm{\Delta_{pa}^2|\mathrm{\Delta_{pc}}}|m\omega_z^2}{g^2\hbar k^2 N}}$, and the superradiant phase has $\mathbb{Z}_n$ or $\mathbb{Z}_{2n}$ symmetry for even or odd $n$, respectively. 

However, upon restoring the cavity-reactive terms, the unbroken-symmetry state becomes unstable. By linearizing Eq.~(\ref{EOMs}) around $(p_j=0,z_j=0)$ and seeking a solution of the form $\tilde{z}_j (t)=\tilde{z}_j e^{-i\omega t}$, we find that for $n>2$, there are two eigenfrequencies given by $\omega^2=\omega_z^2+\frac{N\mathrm{\Omega}^2\hbar k^2g_0^2}{m\mathrm{\Delta_{pa}}^2\sqrt{\mathrm{\Delta_{pc}}^2+\kappa^2}}
\big(\cos\theta\pm i\sin\theta\big)$, where $\tan{\theta}=-\kappa/\mathrm{\Delta_{pc}}$~\cite{methods}. We immediately see that for non-infinite $|\mathrm{\Delta_{pc}}|$ or non-zero $\kappa$, there are eigenfrequencies with positive imaginary part, indicating the presence of exponentially growing eigenmodes. These lead the $z_j=0$ solution to be unstable for any $\mathrm{\Omega}>0$.  

To understand the emergence of steady states at $z_j\neq0$, we focus on the specific case of $n=4$, which admits a graphical solution to the dynamically stationary states.  
In this case, we can reduce the system to two force equations by enforcing the symmetry $z_1=-z_3$ and $z_2=-z_4$. In Fig.\ \ref{fig:M2}(a-c) we plot separately the locus of positions $z_1$ and $z_2$ where the forces $\dot{p}_1$ and $\dot{p}_2$ vanish, restricting our view to the region of position space in which symmetry-broken solutions emerge at the least distance from the cavity nodes. The intersections of these two curves denote stationary states.

Similar models of multiple atomic ensembles that couple to the cavity field with different phases have been shown to possess instabilities and dynamical solutions~\cite{dograDissipationinducedStructuralInstability2019,chiacchioNonreciprocalDickeModel2023,jachinowskiSpinonlyDynamicsMultispecies2025,lyuNonreciprocalGeometricFrustration2025}. In this work, we elucidate additional aspects of the phase diagram with proximity to cavity resonance and pump strength. We identify three dynamical phases: normal, dispersive broken symmetry, and reactive broken symmetry. In the normal phase, which exists at low $\mathrm{\Omega}$, $(z_j=0,\,p_j=0)$ is the only steady state [Fig.~\ref{fig:M2}(a)]. At higher values of $\mathrm{\Omega}$, multiple steady states exist. In the dispersive regime ($|\mathrm{\Delta_{pc}}|\gg\kappa$), there are nine steady states in the region $-\lambda/4<\{z_1,z_2\}<\lambda/4$; four of these steady states break symmetry stably [stars in Fig.~\ref{fig:M2}(b)], while the other five are unstable. We determine the stability by evaluating the eigenvalues of the Jacobian matrix for each steady state~\cite{methods}. Closer to cavity resonance, we enter the reactive regime ($|\mathrm{\Delta_{pc}}|\lesssim \kappa$), where there are no broken symmetry steady states for small values of $|z_1|$ and $|z_2|$, but there are near $z_1,z_2\sim\pm0.5\lambda$ [Fig.~\ref{fig:M2}(c)]; these states are all unstable due to cavity-reactive terms dominating over cavity-dispersive terms in the equations of motion. The transitions between these phases are all discontinuous, with the exception of the transition from normal to broken symmetry in the purely dispersive limit. Numerical analyses performed for $n=3,5,$ and $6$ reveal similar phase diagrams to that shown in Fig.~\ref{fig:M2}(d)~\cite{methods}.

\begin{figure}
    \centering
    \includegraphics{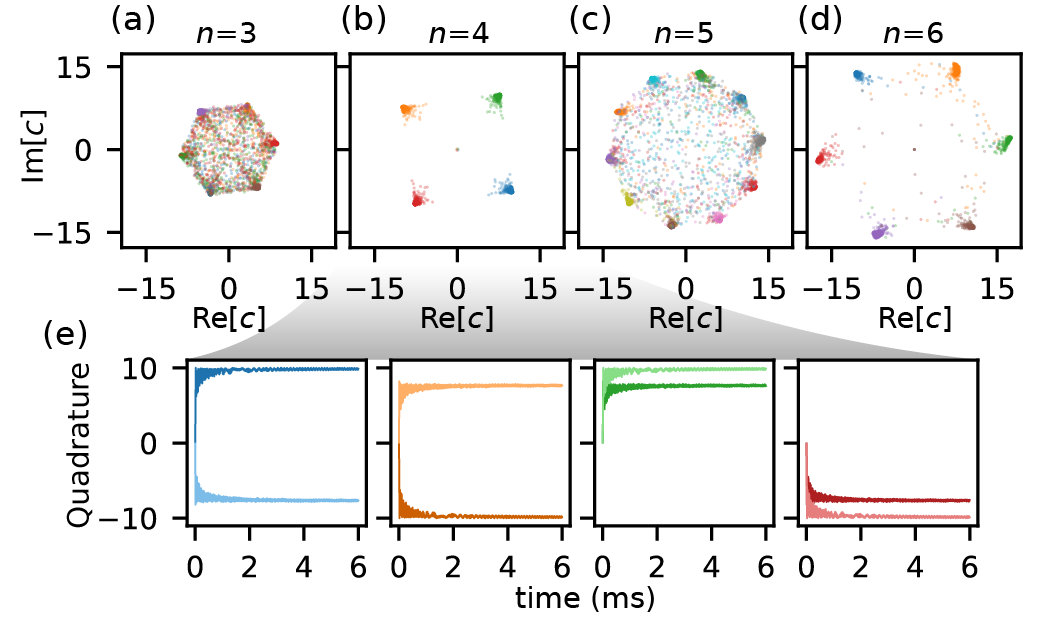}
    \caption{Density plots of cavity field trajectories for (a) $n=3$, (b) $n=4$, (c) $n=5$, and (d) $n=6$ in a parameter regime predicted to have stable steady states. Different colors correspond to different perturbations in the initial positions of the atoms, which are the starting conditions for the numerical integration. The high concentration of single colors at the vertices of a (a) hexagon, (b) square, (c) decagon, and (d) hexagon indicate that the trajectories have spontaneously broken either a $\mathbb{Z}_{2n}$ or a $\mathbb{Z}_n$ symmetry. All trajectories are integrated over a time span of 6~ms. (e) Time trajectories of the real (dark color) and imaginary (light color) quadratures of the cavity field for $n=4$. Colors correspond to the data in (b). Parameters used for all data in figure are $\mathrm{\Omega}=2\pi\times 20$~MHz, $\mathrm{\Delta_{pc}}=-2\pi\times 4$~MHz, $\mathrm{\Delta_{pa}}=-2\pi\times 100$~MHz, $\kappa=2\pi\times 0.5$~MHz, $\nu=30$, $g_0=2\pi\times3$~MHz, and $\omega_z=2\pi\times 70$~kHz.}
    \label{fig:M3}
\end{figure}
Using the same parameters as in Fig.~\ref{fig:M2} while fixing $\mathrm{\Omega}=2\pi\times 20$~MHz and $\mathrm{\Delta_{pc}}=-2\pi\times 4$~MHz, we numerically integrate Eqs.~(\ref{chatdot}-\ref{zhatdot}) to obtain cavity field trajectories for different initial perturbations of the atomic positions and momenta around the $z_j=0$, $p_j=0$ steady state. The results for $n=3,4,5$ and 6 are shown in Fig.~\ref{fig:M3}. The trajectories have a high density at the vertices of an $n$- or $2n$-sided polygon; this confirms the $\mathbb{Z}_{n}$ and $\mathbb{Z}_{2n}$ symmetry breaking as well as the stability of the system's steady states for the chosen parameters. We also examine the time trajectories for $n=4$ [Fig.~\ref{fig:M3}(e)] and see that the cavity field breaks symmetry at late times. We note that the timescale for symmetry breaking appears to shorten when including the dynamics of the atomic excited state, which can also lead to instability when the cavity field is particularly strong~\cite{methods}.

\textit{Description of non-reciprocal interactions---}The distinctive dynamical instabilities of the $n$-phase Dicke model originate from the presence of cavity reactive terms, whose effect grows when $|\mathrm{\Delta_{pc}}|$ approaches $\kappa$.  These terms represent non-reciprocal interactions in this driven-dissipative system. 
To elucidate the role of non-reciprocal forces, we linearize Eq.~(\ref{EOMs}) and rewrite it in terms of bosonic ladder operators $\hat{b}_j$, which annihilate tweezer phonons in the $j$th group of atoms. Defining $\mathbf{b}=(\hat{b}_1,\hat{b}_2,...,\hat{b}_n)^T$, this results in the equation of motion $\dot{\mathbf{b}}=-i\mathbf{H}_\mathrm{eff}\mathbf{b}$ where $\mathbf{H}_\mathrm{eff}$ is an effective Hamiltonian whose matrix elements are given by
\begin{eqnarray}
    \label{H_eff}
    H_{\mathrm{eff},jl}=
    \begin{cases}
        \omega_z+\frac{Ck\mathrm{\Delta_{pc}}}{\mu\omega_z}, \text{ $j=l$}\\
        \frac{Ck\big[\mathrm{\Delta_{pc}}\cos(\phi(j-l))-\kappa\sin(\phi(j-l))\big]}{\mu\omega_z},  \text{ $j\neq l$}
    \end{cases}
\end{eqnarray}
where we have defined $C\equiv\frac{2\hbar\nu^2\mathrm{\Omega}^2}{\mathrm{\Delta_{pa}}^2}\frac{g_0^2k}{{\mathrm{\Delta_{pc}}^2+\kappa^2}}$ for convenience. We see that $H_{\mathrm{eff},jl}\neq H_{\mathrm{eff},lj}$ unless $\kappa\sin(\phi(j-l))=0$ for all values of $j$ and $l$. Since $\phi=2\pi/n$, this results in $\mathbf{H}_\mathrm{eff}$ being non-Hermitian when $n>2$ and $\kappa>0$. The fact that $H_{\mathrm{eff},jl}\neq H_{\mathrm{eff},lj}$ means that the groups of atoms interact non-reciprocally~\cite{ashidaNonHermitianPhysics2020}: the amplitude with which phonons tunnel from group $j$ to group $l$ is not equal to the amplitude with which phonons tunnel from group $l$ to group $j$.

Though the interactions described by $H_\mathrm{eff}$ are generically non-reciprocal, they do not realize ideal non-reciprocal interactions in the sense that it is not possible to have both $H_{\mathrm{eff},jl}=0$ and $H_{\mathrm{eff},lj}\neq0$ for $l\neq j$. However, we now show that one can achieve ideal non-reciprocity if we take $n=2$ and no longer fix the phase difference $\phi$ to be $2\pi/n$. To avoid notational confusion, we define this new, variable phase difference as $\varphi$. The effective Hamiltonian has off-diagonal matrix elements $H_{\mathrm{eff},12}=Ck\big(\mathrm{\Delta_{pc}}\cos\varphi+\kappa\sin\varphi\big)/(\mu\omega_z)$ and $H_{\mathrm{eff},21}=Ck\big(\mathrm{\Delta_{pc}}\cos\varphi-\kappa\sin\varphi\big)/(\mu\omega_z)$.

To illustrate how to achieve a perfectly non-reciprocal interaction, suppose we would like group 1 to be completely decoupled from the motion of group 2, but not vice versa. Achieving this requires $H_{\mathrm{eff},12}=0$, which is satisfied for $\varphi=\tan^{-1}(-\mathrm{\Delta_{pc}}/\kappa)$. We note that this is equivalent to setting $\varphi+\theta=\pm\pi/2$, which has the simple physical interpretation of requiring the cavity field emitted by group 2 to be $\pi/2$ out-of-phase with the pump field at group 1. As such, $\varphi+\theta$ is analogous to a synthetic flux found in other systems with non-reciprocal interactions~\cite{metelmannNonreciprocalPhotonTransmission2015,fangGeneralizedNonreciprocityOptomechanical2017,xuNonreciprocalControlCooling2019}. Such non-reciprocity has been considered in systems of nanoparticles held close together in optical tweezers~\cite{reisenbauerNonHermitianDynamicsNonreciprocity2024a,liskaPTlikePhaseTransition2024,rudolphQuantumOpticalBinding2024,yokomizoNonHermitianPhysicsLevitated2023} and a SiN membrane positioned in an optical cavity~\cite{xuNonreciprocalControlCooling2019}. Here, we have shown that a similar type of interaction can be achieved with trapped atoms in an optical cavity.

\textit{Conclusion---}We have investigated a novel variant of the driven-dissipative Dicke model, in which one can realize $\mathbb{Z}_n$ or $\mathbb{Z}_{2n}$ symmetry breaking by pumping an optomechanical array with $n$ optical phases. For a strong pump, the system stably breaks symmetry in a certain parameter regime, while for a weak pump, it is described by complex eigenfrequencies and non-reciprocal interactions. Our results contribute to the growing literature on instabilities and dynamical solutions in variants of the Dicke model~\cite{dograDissipationinducedStructuralInstability2019,kongkhambutObservationContinuousTime2022,lyuNonreciprocalGeometricFrustration2025,jachinowskiSpinonlyDynamicsMultispecies2025,chiacchioNonreciprocalDickeModel2023}. We have also discussed how to achieve one-way phonon propagation in this system by harnessing control over non-reciprocal interactions. Although we have examined this Dicke model variant in the context of cold atoms, we expect the concepts also to apply to cavity-coupled arrays of other types of mechanical elements, such as thin membranes~\cite{xuNonreciprocalControlCooling2019}, optomechanical crystals~\cite{eichenfieldOptomechanicalCrystals2009}, and optically suspended nanoparticles~\cite{vijayanCavitymediatedLongrangeInteractions2024}. 
One might also consider cavity coupling to the internal spin states of atoms, rather than their mechanical degrees of freedom; engineering of non-reciprocal spin excitation dynamics may have applications in quantum information science~\cite{metelmannNonreciprocalPhotonTransmission2015,xuAutonomousQuantumError2023}. This model additionally has broad prospects for studies of the interplay between symmetries and non-reciprocity in the vicinity of a phase transition~\cite{fruchartNonreciprocalPhaseTransitions2021,zhuNonreciprocalSuperradiantPhase2024,chiacchioNonreciprocalDickeModel2023,zhangNonreciprocalSuperradiantQuantum2025}.

\textit{Acknowledgments---}We thank C. C. Rusconi, S. J. Masson, A. Asenjo-Garcia and N. Vilas for providing review of and insightful comments on this manuscript. We acknowledge support from the AFOSR (Grant No.\ FA9550-1910328), from ARO through the MURI program (Grant No.\ W911NF-20-1-0136), from DARPA (Grant No.\ W911NF2010090), from the NSF (QLCI program through grant number OMA-2016245), and from the U.S. Department of Energy, Office of Science, National Quantum Information Science Research Centers, Quantum Systems Accelerator. J.H. acknowledges support from the Department of Defense through the National Defense Science and Engineering Graduate (NDSEG) Fellowship Program. This work was performed in part at the Aspen Center for Physics, which is supported by National Science Foundation grant PHY-2210452.

\bibliography{references}

\begin{thebibliography}{49}%
\makeatletter
\providecommand \@ifxundefined [1]{%
 \@ifx{#1\undefined}
}%
\providecommand \@ifnum [1]{%
 \ifnum #1\expandafter \@firstoftwo
 \else \expandafter \@secondoftwo
 \fi
}%
\providecommand \@ifx [1]{%
 \ifx #1\expandafter \@firstoftwo
 \else \expandafter \@secondoftwo
 \fi
}%
\providecommand \natexlab [1]{#1}%
\providecommand \enquote  [1]{``#1''}%
\providecommand \bibnamefont  [1]{#1}%
\providecommand \bibfnamefont [1]{#1}%
\providecommand \citenamefont [1]{#1}%
\providecommand \href@noop [0]{\@secondoftwo}%
\providecommand \href [0]{\begingroup \@sanitize@url \@href}%
\providecommand \@href[1]{\@@startlink{#1}\@@href}%
\providecommand \@@href[1]{\endgroup#1\@@endlink}%
\providecommand \@sanitize@url [0]{\catcode `\\12\catcode `\$12\catcode `\&12\catcode `\#12\catcode `\^12\catcode `\_12\catcode `\%12\relax}%
\providecommand \@@startlink[1]{}%
\providecommand \@@endlink[0]{}%
\providecommand \url  [0]{\begingroup\@sanitize@url \@url }%
\providecommand \@url [1]{\endgroup\@href {#1}{\urlprefix }}%
\providecommand \urlprefix  [0]{URL }%
\providecommand \Eprint [0]{\href }%
\providecommand \doibase [0]{https://doi.org/}%
\providecommand \selectlanguage [0]{\@gobble}%
\providecommand \bibinfo  [0]{\@secondoftwo}%
\providecommand \bibfield  [0]{\@secondoftwo}%
\providecommand \translation [1]{[#1]}%
\providecommand \BibitemOpen [0]{}%
\providecommand \bibitemStop [0]{}%
\providecommand \bibitemNoStop [0]{.\EOS\space}%
\providecommand \EOS [0]{\spacefactor3000\relax}%
\providecommand \BibitemShut  [1]{\csname bibitem#1\endcsname}%
\let\auto@bib@innerbib\@empty
\bibitem [{\citenamefont {Hinrichsen}(2006)}]{hinrichsenNonequilibriumPhaseTransitions2006}%
  \BibitemOpen
  \bibfield  {author} {\bibinfo {author} {\bibfnamefont {H.}~\bibnamefont {Hinrichsen}},\ }\bibfield  {title} {\bibinfo {title} {Non-equilibrium phase transitions},\ }\href {https://doi.org/10.1016/j.physa.2006.04.007} {\bibfield  {journal} {\bibinfo  {journal} {Physica A: Statistical Mechanics and its Applications}\ }\textbf {\bibinfo {volume} {369}},\ \bibinfo {pages} {1} (\bibinfo {year} {2006})}\BibitemShut {NoStop}%
\bibitem [{\citenamefont {Matheny}\ \emph {et~al.}(2019)\citenamefont {Matheny}, \citenamefont {Emenheiser}, \citenamefont {Fon}, \citenamefont {Chapman}, \citenamefont {Salova}, \citenamefont {Rohden}, \citenamefont {Li}, \citenamefont {Hudoba De~Badyn}, \citenamefont {P{\'o}sfai}, \citenamefont {{Duenas-Osorio}}, \citenamefont {Mesbahi}, \citenamefont {Crutchfield}, \citenamefont {Cross}, \citenamefont {D'Souza},\ and\ \citenamefont {Roukes}}]{mathenyExoticStatesSimple2019}%
  \BibitemOpen
  \bibfield  {author} {\bibinfo {author} {\bibfnamefont {M.~H.}\ \bibnamefont {Matheny}}, \bibinfo {author} {\bibfnamefont {J.}~\bibnamefont {Emenheiser}}, \bibinfo {author} {\bibfnamefont {W.}~\bibnamefont {Fon}}, \bibinfo {author} {\bibfnamefont {A.}~\bibnamefont {Chapman}}, \bibinfo {author} {\bibfnamefont {A.}~\bibnamefont {Salova}}, \bibinfo {author} {\bibfnamefont {M.}~\bibnamefont {Rohden}}, \bibinfo {author} {\bibfnamefont {J.}~\bibnamefont {Li}}, \bibinfo {author} {\bibfnamefont {M.}~\bibnamefont {Hudoba De~Badyn}}, \bibinfo {author} {\bibfnamefont {M.}~\bibnamefont {P{\'o}sfai}}, \bibinfo {author} {\bibfnamefont {L.}~\bibnamefont {{Duenas-Osorio}}}, \bibinfo {author} {\bibfnamefont {M.}~\bibnamefont {Mesbahi}}, \bibinfo {author} {\bibfnamefont {J.~P.}\ \bibnamefont {Crutchfield}}, \bibinfo {author} {\bibfnamefont {M.~C.}\ \bibnamefont {Cross}}, \bibinfo {author} {\bibfnamefont {R.~M.}\ \bibnamefont {D'Souza}},\ and\ \bibinfo {author} {\bibfnamefont {M.~L.}\ \bibnamefont {Roukes}},\ }\bibfield
  {title} {\bibinfo {title} {Exotic states in a simple network of nanoelectromechanical oscillators},\ }\href {https://doi.org/10.1126/science.aav7932} {\bibfield  {journal} {\bibinfo  {journal} {Science}\ }\textbf {\bibinfo {volume} {363}},\ \bibinfo {pages} {eaav7932} (\bibinfo {year} {2019})}\BibitemShut {NoStop}%
\bibitem [{\citenamefont {Baumann}\ \emph {et~al.}(2011)\citenamefont {Baumann}, \citenamefont {Mottl}, \citenamefont {Brennecke},\ and\ \citenamefont {Esslinger}}]{baumannExploringSymmetryBreaking2011}%
  \BibitemOpen
  \bibfield  {author} {\bibinfo {author} {\bibfnamefont {K.}~\bibnamefont {Baumann}}, \bibinfo {author} {\bibfnamefont {R.}~\bibnamefont {Mottl}}, \bibinfo {author} {\bibfnamefont {F.}~\bibnamefont {Brennecke}},\ and\ \bibinfo {author} {\bibfnamefont {T.}~\bibnamefont {Esslinger}},\ }\bibfield  {title} {\bibinfo {title} {Exploring {{Symmetry Breaking}} at the {{Dicke Quantum Phase Transition}}},\ }\href {https://doi.org/10.1103/PhysRevLett.107.140402} {\bibfield  {journal} {\bibinfo  {journal} {Physical Review Letters}\ }\textbf {\bibinfo {volume} {107}},\ \bibinfo {pages} {140402} (\bibinfo {year} {2011})}\BibitemShut {NoStop}%
\bibitem [{\citenamefont {Young}\ \emph {et~al.}(2020)\citenamefont {Young}, \citenamefont {Gorshkov}, \citenamefont {{Foss-Feig}},\ and\ \citenamefont {Maghrebi}}]{youngNonequilibriumFixedPoints2020}%
  \BibitemOpen
  \bibfield  {author} {\bibinfo {author} {\bibfnamefont {J.~T.}\ \bibnamefont {Young}}, \bibinfo {author} {\bibfnamefont {A.~V.}\ \bibnamefont {Gorshkov}}, \bibinfo {author} {\bibfnamefont {M.}~\bibnamefont {{Foss-Feig}}},\ and\ \bibinfo {author} {\bibfnamefont {M.~F.}\ \bibnamefont {Maghrebi}},\ }\bibfield  {title} {\bibinfo {title} {Nonequilibrium {{Fixed Points}} of {{Coupled Ising Models}}},\ }\href {https://doi.org/10.1103/PhysRevX.10.011039} {\bibfield  {journal} {\bibinfo  {journal} {Physical Review X}\ }\textbf {\bibinfo {volume} {10}},\ \bibinfo {pages} {011039} (\bibinfo {year} {2020})}\BibitemShut {NoStop}%
\bibitem [{\citenamefont {Szyma{\'n}ska}\ \emph {et~al.}(2006)\citenamefont {Szyma{\'n}ska}, \citenamefont {Keeling},\ and\ \citenamefont {Littlewood}}]{szymanskaNonequilibriumQuantumCondensation2006}%
  \BibitemOpen
  \bibfield  {author} {\bibinfo {author} {\bibfnamefont {M.~H.}\ \bibnamefont {Szyma{\'n}ska}}, \bibinfo {author} {\bibfnamefont {J.}~\bibnamefont {Keeling}},\ and\ \bibinfo {author} {\bibfnamefont {P.~B.}\ \bibnamefont {Littlewood}},\ }\bibfield  {title} {\bibinfo {title} {Nonequilibrium {{Quantum Condensation}} in an {{Incoherently Pumped Dissipative System}}},\ }\bibfield  {journal} {\bibinfo  {journal} {Physical Review Letters}\ }\textbf {\bibinfo {volume} {96}},\ \href {https://doi.org/10.1103/physrevlett.96.230602} {10.1103/physrevlett.96.230602} (\bibinfo {year} {2006})\BibitemShut {NoStop}%
\bibitem [{\citenamefont {Belyansky}\ \emph {et~al.}(2025)\citenamefont {Belyansky}, \citenamefont {Weis}, \citenamefont {Hanai}, \citenamefont {Littlewood},\ and\ \citenamefont {Clerk}}]{belyanskyPhaseTransitionsNonreciprocal2025}%
  \BibitemOpen
  \bibfield  {author} {\bibinfo {author} {\bibfnamefont {R.}~\bibnamefont {Belyansky}}, \bibinfo {author} {\bibfnamefont {C.}~\bibnamefont {Weis}}, \bibinfo {author} {\bibfnamefont {R.}~\bibnamefont {Hanai}}, \bibinfo {author} {\bibfnamefont {P.~B.}\ \bibnamefont {Littlewood}},\ and\ \bibinfo {author} {\bibfnamefont {A.~A.}\ \bibnamefont {Clerk}},\ }\bibfield  {title} {\bibinfo {title} {Phase {{Transitions}} in {{Nonreciprocal Driven-Dissipative Condensates}}},\ }\href {https://doi.org/10.1103/gphr-d1bc} {\bibfield  {journal} {\bibinfo  {journal} {Physical Review Letters}\ }\textbf {\bibinfo {volume} {135}},\ \bibinfo {pages} {123401} (\bibinfo {year} {2025})}\BibitemShut {NoStop}%
\bibitem [{\citenamefont {Dicke}(1954)}]{dickeCoherenceSpontaneousRadiation1954}%
  \BibitemOpen
  \bibfield  {author} {\bibinfo {author} {\bibfnamefont {R.~H.}\ \bibnamefont {Dicke}},\ }\bibfield  {title} {\bibinfo {title} {Coherence in {{Spontaneous Radiation Processes}}},\ }\href {https://doi.org/10.1103/PhysRev.93.99} {\bibfield  {journal} {\bibinfo  {journal} {Physical Review}\ }\textbf {\bibinfo {volume} {93}},\ \bibinfo {pages} {99} (\bibinfo {year} {1954})}\BibitemShut {NoStop}%
\bibitem [{\citenamefont {Hepp}\ and\ \citenamefont {Lieb}(1973)}]{heppSuperradiantPhaseTransition1973}%
  \BibitemOpen
  \bibfield  {author} {\bibinfo {author} {\bibfnamefont {K.}~\bibnamefont {Hepp}}\ and\ \bibinfo {author} {\bibfnamefont {E.~H.}\ \bibnamefont {Lieb}},\ }\bibfield  {title} {\bibinfo {title} {On the superradiant phase transition for molecules in a quantized radiation field: The dicke maser model},\ }\href {https://doi.org/10.1016/0003-4916(73)90039-0} {\bibfield  {journal} {\bibinfo  {journal} {Annals of Physics}\ }\textbf {\bibinfo {volume} {76}},\ \bibinfo {pages} {360} (\bibinfo {year} {1973})}\BibitemShut {NoStop}%
\bibitem [{\citenamefont {Sch{\"a}fer}\ \emph {et~al.}(2025)\citenamefont {Sch{\"a}fer}, \citenamefont {Niu}, \citenamefont {Cline}, \citenamefont {Young}, \citenamefont {Song}, \citenamefont {Ritsch},\ and\ \citenamefont {Thompson}}]{schaferContinuousRecoildrivenLasing2025}%
  \BibitemOpen
  \bibfield  {author} {\bibinfo {author} {\bibfnamefont {V.~M.}\ \bibnamefont {Sch{\"a}fer}}, \bibinfo {author} {\bibfnamefont {Z.}~\bibnamefont {Niu}}, \bibinfo {author} {\bibfnamefont {J.~R.~K.}\ \bibnamefont {Cline}}, \bibinfo {author} {\bibfnamefont {D.~J.}\ \bibnamefont {Young}}, \bibinfo {author} {\bibfnamefont {E.~Y.}\ \bibnamefont {Song}}, \bibinfo {author} {\bibfnamefont {H.}~\bibnamefont {Ritsch}},\ and\ \bibinfo {author} {\bibfnamefont {J.~K.}\ \bibnamefont {Thompson}},\ }\bibfield  {title} {\bibinfo {title} {Continuous recoil-driven lasing and cavity frequency pinning with laser-cooled atoms},\ }\href {https://doi.org/10.1038/s41567-025-02854-4} {\bibfield  {journal} {\bibinfo  {journal} {Nature Physics}\ }\textbf {\bibinfo {volume} {21}},\ \bibinfo {pages} {902} (\bibinfo {year} {2025})}\BibitemShut {NoStop}%
\bibitem [{\citenamefont {Norcia}\ \emph {et~al.}(2016)\citenamefont {Norcia}, \citenamefont {Winchester}, \citenamefont {Cline},\ and\ \citenamefont {Thompson}}]{norciaSuperradianceMillihertzLinewidth2016}%
  \BibitemOpen
  \bibfield  {author} {\bibinfo {author} {\bibfnamefont {M.~A.}\ \bibnamefont {Norcia}}, \bibinfo {author} {\bibfnamefont {M.~N.}\ \bibnamefont {Winchester}}, \bibinfo {author} {\bibfnamefont {J.~R.~K.}\ \bibnamefont {Cline}},\ and\ \bibinfo {author} {\bibfnamefont {J.~K.}\ \bibnamefont {Thompson}},\ }\bibfield  {title} {\bibinfo {title} {Superradiance on the millihertz linewidth strontium clock transition},\ }\href {https://doi.org/10.1126/sciadv.1601231} {\bibfield  {journal} {\bibinfo  {journal} {Science Advances}\ }\textbf {\bibinfo {volume} {2}},\ \bibinfo {pages} {e1601231} (\bibinfo {year} {2016})}\BibitemShut {NoStop}%
\bibitem [{\citenamefont {Meiser}\ \emph {et~al.}(2009)\citenamefont {Meiser}, \citenamefont {Ye}, \citenamefont {Carlson},\ and\ \citenamefont {Holland}}]{meiserProspectsMillihertzLinewidthLaser2009}%
  \BibitemOpen
  \bibfield  {author} {\bibinfo {author} {\bibfnamefont {D.}~\bibnamefont {Meiser}}, \bibinfo {author} {\bibfnamefont {J.}~\bibnamefont {Ye}}, \bibinfo {author} {\bibfnamefont {D.~R.}\ \bibnamefont {Carlson}},\ and\ \bibinfo {author} {\bibfnamefont {M.~J.}\ \bibnamefont {Holland}},\ }\bibfield  {title} {\bibinfo {title} {Prospects for a {{Millihertz-Linewidth Laser}}},\ }\href {https://doi.org/10.1103/PhysRevLett.102.163601} {\bibfield  {journal} {\bibinfo  {journal} {Physical Review Letters}\ }\textbf {\bibinfo {volume} {102}},\ \bibinfo {pages} {163601} (\bibinfo {year} {2009})}\BibitemShut {NoStop}%
\bibitem [{\citenamefont {Baumann}\ \emph {et~al.}(2010)\citenamefont {Baumann}, \citenamefont {Guerlin}, \citenamefont {Brennecke},\ and\ \citenamefont {Esslinger}}]{baumannDickeQuantumPhase2010}%
  \BibitemOpen
  \bibfield  {author} {\bibinfo {author} {\bibfnamefont {K.}~\bibnamefont {Baumann}}, \bibinfo {author} {\bibfnamefont {C.}~\bibnamefont {Guerlin}}, \bibinfo {author} {\bibfnamefont {F.}~\bibnamefont {Brennecke}},\ and\ \bibinfo {author} {\bibfnamefont {T.}~\bibnamefont {Esslinger}},\ }\bibfield  {title} {\bibinfo {title} {Dicke quantum phase transition with a superfluid gas in an optical cavity},\ }\href {https://doi.org/10.1038/nature09009} {\bibfield  {journal} {\bibinfo  {journal} {Nature}\ }\textbf {\bibinfo {volume} {464}},\ \bibinfo {pages} {1301} (\bibinfo {year} {2010})}\BibitemShut {NoStop}%
\bibitem [{\citenamefont {Mivehvar}\ \emph {et~al.}(2021)\citenamefont {Mivehvar}, \citenamefont {Piazza}, \citenamefont {Donner},\ and\ \citenamefont {Ritsch}}]{mivehvarCavityQEDQuantum2021}%
  \BibitemOpen
  \bibfield  {author} {\bibinfo {author} {\bibfnamefont {F.}~\bibnamefont {Mivehvar}}, \bibinfo {author} {\bibfnamefont {F.}~\bibnamefont {Piazza}}, \bibinfo {author} {\bibfnamefont {T.}~\bibnamefont {Donner}},\ and\ \bibinfo {author} {\bibfnamefont {H.}~\bibnamefont {Ritsch}},\ }\bibfield  {title} {\bibinfo {title} {Cavity {{QED}} with quantum gases: New paradigms in many-body physics},\ }\href {https://doi.org/10.1080/00018732.2021.1969727} {\bibfield  {journal} {\bibinfo  {journal} {Advances in Physics}\ }\textbf {\bibinfo {volume} {70}},\ \bibinfo {pages} {1} (\bibinfo {year} {2021})}\BibitemShut {NoStop}%
\bibitem [{\citenamefont {Ritsch}\ \emph {et~al.}(2013)\citenamefont {Ritsch}, \citenamefont {Domokos}, \citenamefont {Brennecke},\ and\ \citenamefont {Esslinger}}]{ritschColdAtomsCavitygenerated2013}%
  \BibitemOpen
  \bibfield  {author} {\bibinfo {author} {\bibfnamefont {H.}~\bibnamefont {Ritsch}}, \bibinfo {author} {\bibfnamefont {P.}~\bibnamefont {Domokos}}, \bibinfo {author} {\bibfnamefont {F.}~\bibnamefont {Brennecke}},\ and\ \bibinfo {author} {\bibfnamefont {T.}~\bibnamefont {Esslinger}},\ }\bibfield  {title} {\bibinfo {title} {Cold atoms in cavity-generated dynamical optical potentials},\ }\href {https://doi.org/10.1103/RevModPhys.85.553} {\bibfield  {journal} {\bibinfo  {journal} {Reviews of Modern Physics}\ }\textbf {\bibinfo {volume} {85}},\ \bibinfo {pages} {553} (\bibinfo {year} {2013})}\BibitemShut {NoStop}%
\bibitem [{\citenamefont {{Safavi-Naini}}\ \emph {et~al.}(2018)\citenamefont {{Safavi-Naini}}, \citenamefont {{Lewis-Swan}}, \citenamefont {Bohnet}, \citenamefont {G{\"a}rttner}, \citenamefont {Gilmore}, \citenamefont {Jordan}, \citenamefont {Cohn}, \citenamefont {Freericks}, \citenamefont {Rey},\ and\ \citenamefont {Bollinger}}]{safavi-nainiVerificationManyIonSimulator2018}%
  \BibitemOpen
  \bibfield  {author} {\bibinfo {author} {\bibfnamefont {A.}~\bibnamefont {{Safavi-Naini}}}, \bibinfo {author} {\bibfnamefont {R.~J.}\ \bibnamefont {{Lewis-Swan}}}, \bibinfo {author} {\bibfnamefont {J.~G.}\ \bibnamefont {Bohnet}}, \bibinfo {author} {\bibfnamefont {M.}~\bibnamefont {G{\"a}rttner}}, \bibinfo {author} {\bibfnamefont {K.~A.}\ \bibnamefont {Gilmore}}, \bibinfo {author} {\bibfnamefont {J.~E.}\ \bibnamefont {Jordan}}, \bibinfo {author} {\bibfnamefont {J.}~\bibnamefont {Cohn}}, \bibinfo {author} {\bibfnamefont {J.~K.}\ \bibnamefont {Freericks}}, \bibinfo {author} {\bibfnamefont {A.~M.}\ \bibnamefont {Rey}},\ and\ \bibinfo {author} {\bibfnamefont {J.~J.}\ \bibnamefont {Bollinger}},\ }\bibfield  {title} {\bibinfo {title} {Verification of a {{Many-Ion Simulator}} of the {{Dicke Model Through Slow Quenches}} across a {{Phase Transition}}},\ }\href {https://doi.org/10.1103/PhysRevLett.121.040503} {\bibfield  {journal} {\bibinfo  {journal} {Physical Review Letters}\ }\textbf {\bibinfo {volume} {121}},\
  \bibinfo {pages} {040503} (\bibinfo {year} {2018})}\BibitemShut {NoStop}%
\bibitem [{\citenamefont {Ho}\ \emph {et~al.}(2025)\citenamefont {Ho}, \citenamefont {Lu}, \citenamefont {Xiang}, \citenamefont {Rusconi}, \citenamefont {Masson}, \citenamefont {{Asenjo-Garcia}}, \citenamefont {Yan},\ and\ \citenamefont {{Stamper-Kurn}}}]{hoOptomechanicalSelforganizationMesoscopic2025a}%
  \BibitemOpen
  \bibfield  {author} {\bibinfo {author} {\bibfnamefont {J.}~\bibnamefont {Ho}}, \bibinfo {author} {\bibfnamefont {Y.-H.}\ \bibnamefont {Lu}}, \bibinfo {author} {\bibfnamefont {T.}~\bibnamefont {Xiang}}, \bibinfo {author} {\bibfnamefont {C.~C.}\ \bibnamefont {Rusconi}}, \bibinfo {author} {\bibfnamefont {S.~J.}\ \bibnamefont {Masson}}, \bibinfo {author} {\bibfnamefont {A.}~\bibnamefont {{Asenjo-Garcia}}}, \bibinfo {author} {\bibfnamefont {Z.}~\bibnamefont {Yan}},\ and\ \bibinfo {author} {\bibfnamefont {D.~M.}\ \bibnamefont {{Stamper-Kurn}}},\ }\bibfield  {title} {\bibinfo {title} {Optomechanical self-organization in a mesoscopic atom array},\ }\href {https://doi.org/10.1038/s41567-025-02916-7} {\bibfield  {journal} {\bibinfo  {journal} {Nature Physics}\ }\textbf {\bibinfo {volume} {21}},\ \bibinfo {pages} {1071} (\bibinfo {year} {2025})}\BibitemShut {NoStop}%
\bibitem [{\citenamefont {Zhiqiang}\ \emph {et~al.}(2017)\citenamefont {Zhiqiang}, \citenamefont {Lee}, \citenamefont {Kumar}, \citenamefont {Arnold}, \citenamefont {Masson}, \citenamefont {Parkins},\ and\ \citenamefont {Barrett}}]{zhiqiangNonequilibriumPhaseTransition2017}%
  \BibitemOpen
  \bibfield  {author} {\bibinfo {author} {\bibfnamefont {Z.}~\bibnamefont {Zhiqiang}}, \bibinfo {author} {\bibfnamefont {C.~H.}\ \bibnamefont {Lee}}, \bibinfo {author} {\bibfnamefont {R.}~\bibnamefont {Kumar}}, \bibinfo {author} {\bibfnamefont {K.~J.}\ \bibnamefont {Arnold}}, \bibinfo {author} {\bibfnamefont {S.~J.}\ \bibnamefont {Masson}}, \bibinfo {author} {\bibfnamefont {A.~S.}\ \bibnamefont {Parkins}},\ and\ \bibinfo {author} {\bibfnamefont {M.~D.}\ \bibnamefont {Barrett}},\ }\bibfield  {title} {\bibinfo {title} {Nonequilibrium phase transition in a spin-1 {{Dicke}} model},\ }\href {https://doi.org/10.1364/OPTICA.4.000424} {\bibfield  {journal} {\bibinfo  {journal} {Optica}\ }\textbf {\bibinfo {volume} {4}},\ \bibinfo {pages} {424} (\bibinfo {year} {2017})}\BibitemShut {NoStop}%
\bibitem [{\citenamefont {Nairn}\ \emph {et~al.}(2025)\citenamefont {Nairn}, \citenamefont {Giannelli}, \citenamefont {Morigi}, \citenamefont {Slama}, \citenamefont {Olmos},\ and\ \citenamefont {J{\"a}ger}}]{nairnSpinSelfOrganizationOptical2025}%
  \BibitemOpen
  \bibfield  {author} {\bibinfo {author} {\bibfnamefont {M.}~\bibnamefont {Nairn}}, \bibinfo {author} {\bibfnamefont {L.}~\bibnamefont {Giannelli}}, \bibinfo {author} {\bibfnamefont {G.}~\bibnamefont {Morigi}}, \bibinfo {author} {\bibfnamefont {S.}~\bibnamefont {Slama}}, \bibinfo {author} {\bibfnamefont {B.}~\bibnamefont {Olmos}},\ and\ \bibinfo {author} {\bibfnamefont {S.~B.}\ \bibnamefont {J{\"a}ger}},\ }\bibfield  {title} {\bibinfo {title} {Spin {{Self-Organization}} in an {{Optical Cavity Facilitated}} by {{Inhomogeneous Broadening}}},\ }\href {https://doi.org/10.1103/PhysRevLett.134.083603} {\bibfield  {journal} {\bibinfo  {journal} {Physical Review Letters}\ }\textbf {\bibinfo {volume} {134}},\ \bibinfo {pages} {083603} (\bibinfo {year} {2025})}\BibitemShut {NoStop}%
\bibitem [{\citenamefont {Temnov}\ and\ \citenamefont {Woggon}(2005)}]{temnovSuperradianceSubradianceInhomogeneously2005}%
  \BibitemOpen
  \bibfield  {author} {\bibinfo {author} {\bibfnamefont {V.~V.}\ \bibnamefont {Temnov}}\ and\ \bibinfo {author} {\bibfnamefont {U.}~\bibnamefont {Woggon}},\ }\bibfield  {title} {\bibinfo {title} {Superradiance and {{Subradiance}} in an {{Inhomogeneously Broadened Ensemble}} of {{Two-Level Systems Coupled}} to a {{Low- Q Cavity}}},\ }\href {https://doi.org/10.1103/PhysRevLett.95.243602} {\bibfield  {journal} {\bibinfo  {journal} {Physical Review Letters}\ }\textbf {\bibinfo {volume} {95}},\ \bibinfo {pages} {243602} (\bibinfo {year} {2005})}\BibitemShut {NoStop}%
\bibitem [{\citenamefont {Das}\ \emph {et~al.}(2024)\citenamefont {Das}, \citenamefont {W{\"u}ster},\ and\ \citenamefont {Sharma}}]{dasDickeModelDisordered2024}%
  \BibitemOpen
  \bibfield  {author} {\bibinfo {author} {\bibfnamefont {P.}~\bibnamefont {Das}}, \bibinfo {author} {\bibfnamefont {S.}~\bibnamefont {W{\"u}ster}},\ and\ \bibinfo {author} {\bibfnamefont {A.}~\bibnamefont {Sharma}},\ }\bibfield  {title} {\bibinfo {title} {Dicke model with disordered spin-boson couplings},\ }\href {https://doi.org/10.1103/PhysRevA.109.013715} {\bibfield  {journal} {\bibinfo  {journal} {Physical Review A}\ }\textbf {\bibinfo {volume} {109}},\ \bibinfo {pages} {013715} (\bibinfo {year} {2024})}\BibitemShut {NoStop}%
\bibitem [{\citenamefont {Zhang}\ \emph {et~al.}(2018)\citenamefont {Zhang}, \citenamefont {Lee}, \citenamefont {Kumar}, \citenamefont {Arnold}, \citenamefont {Masson}, \citenamefont {Grimsmo}, \citenamefont {Parkins},\ and\ \citenamefont {Barrett}}]{zhangDickemodelSimulationCavityassisted2018}%
  \BibitemOpen
  \bibfield  {author} {\bibinfo {author} {\bibfnamefont {Z.}~\bibnamefont {Zhang}}, \bibinfo {author} {\bibfnamefont {C.~H.}\ \bibnamefont {Lee}}, \bibinfo {author} {\bibfnamefont {R.}~\bibnamefont {Kumar}}, \bibinfo {author} {\bibfnamefont {K.~J.}\ \bibnamefont {Arnold}}, \bibinfo {author} {\bibfnamefont {S.~J.}\ \bibnamefont {Masson}}, \bibinfo {author} {\bibfnamefont {A.~L.}\ \bibnamefont {Grimsmo}}, \bibinfo {author} {\bibfnamefont {A.~S.}\ \bibnamefont {Parkins}},\ and\ \bibinfo {author} {\bibfnamefont {M.~D.}\ \bibnamefont {Barrett}},\ }\bibfield  {title} {\bibinfo {title} {Dicke-model simulation via cavity-assisted {{Raman}} transitions},\ }\href {https://doi.org/10.1103/PhysRevA.97.043858} {\bibfield  {journal} {\bibinfo  {journal} {Physical Review A}\ }\textbf {\bibinfo {volume} {97}},\ \bibinfo {pages} {043858} (\bibinfo {year} {2018})}\BibitemShut {NoStop}%
\bibitem [{\citenamefont {Kroeze}\ \emph {et~al.}(2025)\citenamefont {Kroeze}, \citenamefont {Marsh}, \citenamefont {Schuller}, \citenamefont {Hunt}, \citenamefont {Bourzutschky}, \citenamefont {Winer}, \citenamefont {Gopalakrishnan}, \citenamefont {Keeling},\ and\ \citenamefont {Lev}}]{kroezeDirectlyObservingReplica2025}%
  \BibitemOpen
  \bibfield  {author} {\bibinfo {author} {\bibfnamefont {R.~M.}\ \bibnamefont {Kroeze}}, \bibinfo {author} {\bibfnamefont {B.~P.}\ \bibnamefont {Marsh}}, \bibinfo {author} {\bibfnamefont {D.~A.}\ \bibnamefont {Schuller}}, \bibinfo {author} {\bibfnamefont {H.~S.}\ \bibnamefont {Hunt}}, \bibinfo {author} {\bibfnamefont {A.~N.}\ \bibnamefont {Bourzutschky}}, \bibinfo {author} {\bibfnamefont {M.}~\bibnamefont {Winer}}, \bibinfo {author} {\bibfnamefont {S.}~\bibnamefont {Gopalakrishnan}}, \bibinfo {author} {\bibfnamefont {J.}~\bibnamefont {Keeling}},\ and\ \bibinfo {author} {\bibfnamefont {B.~L.}\ \bibnamefont {Lev}},\ }\bibfield  {title} {\bibinfo {title} {Directly observing replica symmetry breaking in a vector quantum-optical spin glass},\ }\href {https://doi.org/10.1126/science.adu7710} {\bibfield  {journal} {\bibinfo  {journal} {Science}\ }\textbf {\bibinfo {volume} {389}},\ \bibinfo {pages} {1122} (\bibinfo {year} {2025})}\BibitemShut {NoStop}%
\bibitem [{\citenamefont {Marsh}\ \emph {et~al.}(2024)\citenamefont {Marsh}, \citenamefont {Kroeze}, \citenamefont {Ganguli}, \citenamefont {Gopalakrishnan}, \citenamefont {Keeling},\ and\ \citenamefont {Lev}}]{marshEntanglementReplicaSymmetry2024}%
  \BibitemOpen
  \bibfield  {author} {\bibinfo {author} {\bibfnamefont {B.~P.}\ \bibnamefont {Marsh}}, \bibinfo {author} {\bibfnamefont {R.~M.}\ \bibnamefont {Kroeze}}, \bibinfo {author} {\bibfnamefont {S.}~\bibnamefont {Ganguli}}, \bibinfo {author} {\bibfnamefont {S.}~\bibnamefont {Gopalakrishnan}}, \bibinfo {author} {\bibfnamefont {J.}~\bibnamefont {Keeling}},\ and\ \bibinfo {author} {\bibfnamefont {B.~L.}\ \bibnamefont {Lev}},\ }\bibfield  {title} {\bibinfo {title} {Entanglement and {{Replica Symmetry Breaking}} in a {{Driven-Dissipative Quantum Spin Glass}}},\ }\href {https://doi.org/10.1103/PhysRevX.14.011026} {\bibfield  {journal} {\bibinfo  {journal} {Physical Review X}\ }\textbf {\bibinfo {volume} {14}},\ \bibinfo {pages} {011026} (\bibinfo {year} {2024})}\BibitemShut {NoStop}%
\bibitem [{\citenamefont {Wu}(1982)}]{wuPottsModel1982}%
  \BibitemOpen
  \bibfield  {author} {\bibinfo {author} {\bibfnamefont {F.~Y.}\ \bibnamefont {Wu}},\ }\bibfield  {title} {\bibinfo {title} {The {{Potts}} model},\ }\href {https://doi.org/10.1103/RevModPhys.54.235} {\bibfield  {journal} {\bibinfo  {journal} {Reviews of Modern Physics}\ }\textbf {\bibinfo {volume} {54}},\ \bibinfo {pages} {235} (\bibinfo {year} {1982})}\BibitemShut {NoStop}%
\bibitem [{\citenamefont {Chatterjee}\ \emph {et~al.}(2018)\citenamefont {Chatterjee}, \citenamefont {Puri},\ and\ \citenamefont {Paul}}]{chatterjeeOrderingKineticsState2018}%
  \BibitemOpen
  \bibfield  {author} {\bibinfo {author} {\bibfnamefont {S.}~\bibnamefont {Chatterjee}}, \bibinfo {author} {\bibfnamefont {S.}~\bibnamefont {Puri}},\ and\ \bibinfo {author} {\bibfnamefont {R.}~\bibnamefont {Paul}},\ }\bibfield  {title} {\bibinfo {title} {Ordering kinetics in the q -state clock model: {{Scaling}} properties and growth laws},\ }\href {https://doi.org/10.1103/PhysRevE.98.032109} {\bibfield  {journal} {\bibinfo  {journal} {Physical Review E}\ }\textbf {\bibinfo {volume} {98}},\ \bibinfo {pages} {032109} (\bibinfo {year} {2018})}\BibitemShut {NoStop}%
\bibitem [{\citenamefont {Wang}\ \emph {et~al.}(2025)\citenamefont {Wang}, \citenamefont {Spierings}, \citenamefont {Peters}, \citenamefont {Chen}, \citenamefont {Deli{\'c}},\ and\ \citenamefont {Vuleti{\'c}}}]{wangProgrammableFewatomBragg2025}%
  \BibitemOpen
  \bibfield  {author} {\bibinfo {author} {\bibfnamefont {G.}~\bibnamefont {Wang}}, \bibinfo {author} {\bibfnamefont {D.~C.}\ \bibnamefont {Spierings}}, \bibinfo {author} {\bibfnamefont {M.~L.}\ \bibnamefont {Peters}}, \bibinfo {author} {\bibfnamefont {M.-W.}\ \bibnamefont {Chen}}, \bibinfo {author} {\bibfnamefont {U.}~\bibnamefont {Deli{\'c}}},\ and\ \bibinfo {author} {\bibfnamefont {V.}~\bibnamefont {Vuleti{\'c}}},\ }\href {https://doi.org/10.48550/arXiv.2508.10748} {\bibinfo {title} {Programmable few-atom {{Bragg}} scattering and ground-state cooling in a cavity}} (\bibinfo {year} {2025}),\ \Eprint {https://arxiv.org/abs/2508.10748} {arXiv:2508.10748 [quant-ph]} \BibitemShut {NoStop}%
\bibitem [{\citenamefont {Holstein}\ and\ \citenamefont {Primakoff}(1940)}]{holsteinFieldDependenceIntrinsic1940}%
  \BibitemOpen
  \bibfield  {author} {\bibinfo {author} {\bibfnamefont {T.}~\bibnamefont {Holstein}}\ and\ \bibinfo {author} {\bibfnamefont {H.}~\bibnamefont {Primakoff}},\ }\bibfield  {title} {\bibinfo {title} {Field {{Dependence}} of the {{Intrinsic Domain Magnetization}} of a {{Ferromagnet}}},\ }\href {https://doi.org/10.1103/PhysRev.58.1098} {\bibfield  {journal} {\bibinfo  {journal} {Physical Review}\ }\textbf {\bibinfo {volume} {58}},\ \bibinfo {pages} {1098} (\bibinfo {year} {1940})}\BibitemShut {NoStop}%
\bibitem [{met()}]{methods}%
  \BibitemOpen
  \href@noop {} {}\bibinfo {note} {See Supplemental Material.}\BibitemShut {Stop}%
\bibitem [{\citenamefont {Schaeffer}\ and\ \citenamefont {Cain}(2016)}]{schaefferOrdinaryDifferentialEquations2016}%
  \BibitemOpen
  \bibfield  {author} {\bibinfo {author} {\bibfnamefont {D.~G.}\ \bibnamefont {Schaeffer}}\ and\ \bibinfo {author} {\bibfnamefont {J.~W.}\ \bibnamefont {Cain}},\ }\href {https://doi.org/10.1007/978-1-4939-6389-8} {\emph {\bibinfo {title} {Ordinary {{Differential Equations}}: {{Basics}} and {{Beyond}}}}},\ Texts in {{Applied Mathematics}}\ (\bibinfo  {publisher} {Springer New York},\ \bibinfo {address} {New York, NY},\ \bibinfo {year} {2016})\BibitemShut {NoStop}%
\bibitem [{\citenamefont {Endres}\ \emph {et~al.}(2018)\citenamefont {Endres}, \citenamefont {Sandrock},\ and\ \citenamefont {Focke}}]{endresSimplicialHomologyAlgorithm2018}%
  \BibitemOpen
  \bibfield  {author} {\bibinfo {author} {\bibfnamefont {S.~C.}\ \bibnamefont {Endres}}, \bibinfo {author} {\bibfnamefont {C.}~\bibnamefont {Sandrock}},\ and\ \bibinfo {author} {\bibfnamefont {W.~W.}\ \bibnamefont {Focke}},\ }\bibfield  {title} {\bibinfo {title} {A simplicial homology algorithm for {{Lipschitz}} optimisation},\ }\href {https://doi.org/10.1007/s10898-018-0645-y} {\bibfield  {journal} {\bibinfo  {journal} {Journal of Global Optimization}\ }\textbf {\bibinfo {volume} {72}},\ \bibinfo {pages} {181} (\bibinfo {year} {2018})}\BibitemShut {NoStop}%
\bibitem [{\citenamefont {Dogra}\ \emph {et~al.}(2019)\citenamefont {Dogra}, \citenamefont {Landini}, \citenamefont {Kroeger}, \citenamefont {Hruby}, \citenamefont {Donner},\ and\ \citenamefont {Esslinger}}]{dograDissipationinducedStructuralInstability2019}%
  \BibitemOpen
  \bibfield  {author} {\bibinfo {author} {\bibfnamefont {N.}~\bibnamefont {Dogra}}, \bibinfo {author} {\bibfnamefont {M.}~\bibnamefont {Landini}}, \bibinfo {author} {\bibfnamefont {K.}~\bibnamefont {Kroeger}}, \bibinfo {author} {\bibfnamefont {L.}~\bibnamefont {Hruby}}, \bibinfo {author} {\bibfnamefont {T.}~\bibnamefont {Donner}},\ and\ \bibinfo {author} {\bibfnamefont {T.}~\bibnamefont {Esslinger}},\ }\bibfield  {title} {\bibinfo {title} {Dissipation-induced structural instability and chiral dynamics in a quantum gas},\ }\href {https://doi.org/10.1126/science.aaw4465} {\bibfield  {journal} {\bibinfo  {journal} {Science}\ }\textbf {\bibinfo {volume} {366}},\ \bibinfo {pages} {1496} (\bibinfo {year} {2019})}\BibitemShut {NoStop}%
\bibitem [{\citenamefont {Chiacchio}\ \emph {et~al.}(2023)\citenamefont {Chiacchio}, \citenamefont {Nunnenkamp},\ and\ \citenamefont {Brunelli}}]{chiacchioNonreciprocalDickeModel2023}%
  \BibitemOpen
  \bibfield  {author} {\bibinfo {author} {\bibfnamefont {E.~I.~R.}\ \bibnamefont {Chiacchio}}, \bibinfo {author} {\bibfnamefont {A.}~\bibnamefont {Nunnenkamp}},\ and\ \bibinfo {author} {\bibfnamefont {M.}~\bibnamefont {Brunelli}},\ }\bibfield  {title} {\bibinfo {title} {Nonreciprocal {{Dicke Model}}},\ }\href {https://doi.org/10.1103/PhysRevLett.131.113602} {\bibfield  {journal} {\bibinfo  {journal} {Physical Review Letters}\ }\textbf {\bibinfo {volume} {131}},\ \bibinfo {pages} {113602} (\bibinfo {year} {2023})}\BibitemShut {NoStop}%
\bibitem [{\citenamefont {Jachinowski}\ and\ \citenamefont {Littlewood}(2025)}]{jachinowskiSpinonlyDynamicsMultispecies2025}%
  \BibitemOpen
  \bibfield  {author} {\bibinfo {author} {\bibfnamefont {J.}~\bibnamefont {Jachinowski}}\ and\ \bibinfo {author} {\bibfnamefont {P.~B.}\ \bibnamefont {Littlewood}},\ }\href {https://doi.org/10.48550/arXiv.2507.07960} {\bibinfo {title} {Spin-only dynamics of the multi-species nonreciprocal {{Dicke}} model}} (\bibinfo {year} {2025}),\ \Eprint {https://arxiv.org/abs/2507.07960} {arXiv:2507.07960 [cond-mat]} \BibitemShut {NoStop}%
\bibitem [{\citenamefont {Lyu}\ and\ \citenamefont {Hwang}(2025)}]{lyuNonreciprocalGeometricFrustration2025}%
  \BibitemOpen
  \bibfield  {author} {\bibinfo {author} {\bibfnamefont {G.}~\bibnamefont {Lyu}}\ and\ \bibinfo {author} {\bibfnamefont {M.-J.}\ \bibnamefont {Hwang}},\ }\href {https://doi.org/10.48550/arXiv.2508.06444} {\bibinfo {title} {Nonreciprocal and {{Geometric Frustration}} in {{Dissipative Quantum Spins}}}} (\bibinfo {year} {2025}),\ \Eprint {https://arxiv.org/abs/2508.06444} {arXiv:2508.06444 [quant-ph]} \BibitemShut {NoStop}%
\bibitem [{\citenamefont {Ashida}\ \emph {et~al.}(2020)\citenamefont {Ashida}, \citenamefont {Gong},\ and\ \citenamefont {Ueda}}]{ashidaNonHermitianPhysics2020}%
  \BibitemOpen
  \bibfield  {author} {\bibinfo {author} {\bibfnamefont {Y.}~\bibnamefont {Ashida}}, \bibinfo {author} {\bibfnamefont {Z.}~\bibnamefont {Gong}},\ and\ \bibinfo {author} {\bibfnamefont {M.}~\bibnamefont {Ueda}},\ }\bibfield  {title} {\bibinfo {title} {Non-{{Hermitian}} physics},\ }\href {https://doi.org/10.1080/00018732.2021.1876991} {\bibfield  {journal} {\bibinfo  {journal} {Advances in Physics}\ }\textbf {\bibinfo {volume} {69}},\ \bibinfo {pages} {249} (\bibinfo {year} {2020})}\BibitemShut {NoStop}%
\bibitem [{\citenamefont {Metelmann}\ and\ \citenamefont {Clerk}(2015)}]{metelmannNonreciprocalPhotonTransmission2015}%
  \BibitemOpen
  \bibfield  {author} {\bibinfo {author} {\bibfnamefont {A.}~\bibnamefont {Metelmann}}\ and\ \bibinfo {author} {\bibfnamefont {A.~A.}\ \bibnamefont {Clerk}},\ }\bibfield  {title} {\bibinfo {title} {Nonreciprocal {{Photon Transmission}} and {{Amplification}} via {{Reservoir Engineering}}},\ }\href {https://doi.org/10.1103/PhysRevX.5.021025} {\bibfield  {journal} {\bibinfo  {journal} {Physical Review X}\ }\textbf {\bibinfo {volume} {5}},\ \bibinfo {pages} {021025} (\bibinfo {year} {2015})}\BibitemShut {NoStop}%
\bibitem [{\citenamefont {Fang}\ \emph {et~al.}(2017)\citenamefont {Fang}, \citenamefont {Luo}, \citenamefont {Metelmann}, \citenamefont {Matheny}, \citenamefont {Marquardt}, \citenamefont {Clerk},\ and\ \citenamefont {Painter}}]{fangGeneralizedNonreciprocityOptomechanical2017}%
  \BibitemOpen
  \bibfield  {author} {\bibinfo {author} {\bibfnamefont {K.}~\bibnamefont {Fang}}, \bibinfo {author} {\bibfnamefont {J.}~\bibnamefont {Luo}}, \bibinfo {author} {\bibfnamefont {A.}~\bibnamefont {Metelmann}}, \bibinfo {author} {\bibfnamefont {M.~H.}\ \bibnamefont {Matheny}}, \bibinfo {author} {\bibfnamefont {F.}~\bibnamefont {Marquardt}}, \bibinfo {author} {\bibfnamefont {A.~A.}\ \bibnamefont {Clerk}},\ and\ \bibinfo {author} {\bibfnamefont {O.}~\bibnamefont {Painter}},\ }\bibfield  {title} {\bibinfo {title} {Generalized non-reciprocity in an optomechanical circuit via synthetic magnetism and reservoir engineering},\ }\href {https://doi.org/10.1038/nphys4009} {\bibfield  {journal} {\bibinfo  {journal} {Nature Physics}\ }\textbf {\bibinfo {volume} {13}},\ \bibinfo {pages} {465} (\bibinfo {year} {2017})}\BibitemShut {NoStop}%
\bibitem [{\citenamefont {Xu}\ \emph {et~al.}(2019)\citenamefont {Xu}, \citenamefont {Jiang}, \citenamefont {Clerk},\ and\ \citenamefont {Harris}}]{xuNonreciprocalControlCooling2019}%
  \BibitemOpen
  \bibfield  {author} {\bibinfo {author} {\bibfnamefont {H.}~\bibnamefont {Xu}}, \bibinfo {author} {\bibfnamefont {L.}~\bibnamefont {Jiang}}, \bibinfo {author} {\bibfnamefont {A.~A.}\ \bibnamefont {Clerk}},\ and\ \bibinfo {author} {\bibfnamefont {J.~G.~E.}\ \bibnamefont {Harris}},\ }\bibfield  {title} {\bibinfo {title} {Nonreciprocal control and cooling of phonon modes in an optomechanical system},\ }\href {https://doi.org/10.1038/s41586-019-1061-2} {\bibfield  {journal} {\bibinfo  {journal} {Nature}\ }\textbf {\bibinfo {volume} {568}},\ \bibinfo {pages} {65} (\bibinfo {year} {2019})}\BibitemShut {NoStop}%
\bibitem [{\citenamefont {Reisenbauer}\ \emph {et~al.}(2024)\citenamefont {Reisenbauer}, \citenamefont {Rudolph}, \citenamefont {Egyed}, \citenamefont {Hornberger}, \citenamefont {Zasedatelev}, \citenamefont {Abuzarli}, \citenamefont {Stickler},\ and\ \citenamefont {Deli{\'c}}}]{reisenbauerNonHermitianDynamicsNonreciprocity2024a}%
  \BibitemOpen
  \bibfield  {author} {\bibinfo {author} {\bibfnamefont {M.}~\bibnamefont {Reisenbauer}}, \bibinfo {author} {\bibfnamefont {H.}~\bibnamefont {Rudolph}}, \bibinfo {author} {\bibfnamefont {L.}~\bibnamefont {Egyed}}, \bibinfo {author} {\bibfnamefont {K.}~\bibnamefont {Hornberger}}, \bibinfo {author} {\bibfnamefont {A.~V.}\ \bibnamefont {Zasedatelev}}, \bibinfo {author} {\bibfnamefont {M.}~\bibnamefont {Abuzarli}}, \bibinfo {author} {\bibfnamefont {B.~A.}\ \bibnamefont {Stickler}},\ and\ \bibinfo {author} {\bibfnamefont {U.}~\bibnamefont {Deli{\'c}}},\ }\bibfield  {title} {\bibinfo {title} {Non-{{Hermitian}} dynamics and non-reciprocity of optically coupled nanoparticles},\ }\href {https://doi.org/10.1038/s41567-024-02589-8} {\bibfield  {journal} {\bibinfo  {journal} {Nature Physics}\ }\textbf {\bibinfo {volume} {20}},\ \bibinfo {pages} {1629} (\bibinfo {year} {2024})}\BibitemShut {NoStop}%
\bibitem [{\citenamefont {Li{\v s}ka}\ \emph {et~al.}(2024)\citenamefont {Li{\v s}ka}, \citenamefont {Zem{\'a}nkov{\'a}}, \citenamefont {J{\'a}kl}, \citenamefont {{\v S}iler}, \citenamefont {Simpson}, \citenamefont {Zem{\'a}nek},\ and\ \citenamefont {Brzobohat{\'y}}}]{liskaPTlikePhaseTransition2024}%
  \BibitemOpen
  \bibfield  {author} {\bibinfo {author} {\bibfnamefont {V.}~\bibnamefont {Li{\v s}ka}}, \bibinfo {author} {\bibfnamefont {T.}~\bibnamefont {Zem{\'a}nkov{\'a}}}, \bibinfo {author} {\bibfnamefont {P.}~\bibnamefont {J{\'a}kl}}, \bibinfo {author} {\bibfnamefont {M.}~\bibnamefont {{\v S}iler}}, \bibinfo {author} {\bibfnamefont {S.~H.}\ \bibnamefont {Simpson}}, \bibinfo {author} {\bibfnamefont {P.}~\bibnamefont {Zem{\'a}nek}},\ and\ \bibinfo {author} {\bibfnamefont {O.}~\bibnamefont {Brzobohat{\'y}}},\ }\bibfield  {title} {\bibinfo {title} {{{PT-like}} phase transition and limit cycle oscillations in non-reciprocally coupled optomechanical oscillators levitated in vacuum},\ }\href {https://doi.org/10.1038/s41567-024-02590-1} {\bibfield  {journal} {\bibinfo  {journal} {Nature Physics}\ }\textbf {\bibinfo {volume} {20}},\ \bibinfo {pages} {1622} (\bibinfo {year} {2024})}\BibitemShut {NoStop}%
\bibitem [{\citenamefont {Rudolph}\ \emph {et~al.}(2024)\citenamefont {Rudolph}, \citenamefont {Deli{\'c}}, \citenamefont {Hornberger},\ and\ \citenamefont {Stickler}}]{rudolphQuantumOpticalBinding2024}%
  \BibitemOpen
  \bibfield  {author} {\bibinfo {author} {\bibfnamefont {H.}~\bibnamefont {Rudolph}}, \bibinfo {author} {\bibfnamefont {U.}~\bibnamefont {Deli{\'c}}}, \bibinfo {author} {\bibfnamefont {K.}~\bibnamefont {Hornberger}},\ and\ \bibinfo {author} {\bibfnamefont {B.~A.}\ \bibnamefont {Stickler}},\ }\bibfield  {title} {\bibinfo {title} {Quantum {{Optical Binding}} of {{Nanoscale Particles}}},\ }\href {https://doi.org/10.1103/PhysRevLett.133.233603} {\bibfield  {journal} {\bibinfo  {journal} {Physical Review Letters}\ }\textbf {\bibinfo {volume} {133}},\ \bibinfo {pages} {233603} (\bibinfo {year} {2024})}\BibitemShut {NoStop}%
\bibitem [{\citenamefont {Yokomizo}\ and\ \citenamefont {Ashida}(2023)}]{yokomizoNonHermitianPhysicsLevitated2023}%
  \BibitemOpen
  \bibfield  {author} {\bibinfo {author} {\bibfnamefont {K.}~\bibnamefont {Yokomizo}}\ and\ \bibinfo {author} {\bibfnamefont {Y.}~\bibnamefont {Ashida}},\ }\bibfield  {title} {\bibinfo {title} {Non-{{Hermitian}} physics of levitated nanoparticle array},\ }\href {https://doi.org/10.1103/PhysRevResearch.5.033217} {\bibfield  {journal} {\bibinfo  {journal} {Physical Review Research}\ }\textbf {\bibinfo {volume} {5}},\ \bibinfo {pages} {033217} (\bibinfo {year} {2023})}\BibitemShut {NoStop}%
\bibitem [{\citenamefont {Kongkhambut}\ \emph {et~al.}(2022)\citenamefont {Kongkhambut}, \citenamefont {Skulte}, \citenamefont {Mathey}, \citenamefont {Cosme}, \citenamefont {Hemmerich},\ and\ \citenamefont {Ke{\ss}ler}}]{kongkhambutObservationContinuousTime2022}%
  \BibitemOpen
  \bibfield  {author} {\bibinfo {author} {\bibfnamefont {P.}~\bibnamefont {Kongkhambut}}, \bibinfo {author} {\bibfnamefont {J.}~\bibnamefont {Skulte}}, \bibinfo {author} {\bibfnamefont {L.}~\bibnamefont {Mathey}}, \bibinfo {author} {\bibfnamefont {J.~G.}\ \bibnamefont {Cosme}}, \bibinfo {author} {\bibfnamefont {A.}~\bibnamefont {Hemmerich}},\ and\ \bibinfo {author} {\bibfnamefont {H.}~\bibnamefont {Ke{\ss}ler}},\ }\bibfield  {title} {\bibinfo {title} {Observation of a continuous time crystal},\ }\href {https://doi.org/10.1126/science.abo3382} {\bibfield  {journal} {\bibinfo  {journal} {Science}\ }\textbf {\bibinfo {volume} {377}},\ \bibinfo {pages} {670} (\bibinfo {year} {2022})}\BibitemShut {NoStop}%
\bibitem [{\citenamefont {Eichenfield}\ \emph {et~al.}(2009)\citenamefont {Eichenfield}, \citenamefont {Chan}, \citenamefont {Camacho}, \citenamefont {Vahala},\ and\ \citenamefont {Painter}}]{eichenfieldOptomechanicalCrystals2009}%
  \BibitemOpen
  \bibfield  {author} {\bibinfo {author} {\bibfnamefont {M.}~\bibnamefont {Eichenfield}}, \bibinfo {author} {\bibfnamefont {J.}~\bibnamefont {Chan}}, \bibinfo {author} {\bibfnamefont {R.~M.}\ \bibnamefont {Camacho}}, \bibinfo {author} {\bibfnamefont {K.~J.}\ \bibnamefont {Vahala}},\ and\ \bibinfo {author} {\bibfnamefont {O.}~\bibnamefont {Painter}},\ }\bibfield  {title} {\bibinfo {title} {Optomechanical crystals},\ }\href {https://doi.org/10.1038/nature08524} {\bibfield  {journal} {\bibinfo  {journal} {Nature}\ }\textbf {\bibinfo {volume} {462}},\ \bibinfo {pages} {78} (\bibinfo {year} {2009})}\BibitemShut {NoStop}%
\bibitem [{\citenamefont {Vijayan}\ \emph {et~al.}(2024)\citenamefont {Vijayan}, \citenamefont {Piotrowski}, \citenamefont {{Gonzalez-Ballestero}}, \citenamefont {Weber}, \citenamefont {{Romero-Isart}},\ and\ \citenamefont {Novotny}}]{vijayanCavitymediatedLongrangeInteractions2024}%
  \BibitemOpen
  \bibfield  {author} {\bibinfo {author} {\bibfnamefont {J.}~\bibnamefont {Vijayan}}, \bibinfo {author} {\bibfnamefont {J.}~\bibnamefont {Piotrowski}}, \bibinfo {author} {\bibfnamefont {C.}~\bibnamefont {{Gonzalez-Ballestero}}}, \bibinfo {author} {\bibfnamefont {K.}~\bibnamefont {Weber}}, \bibinfo {author} {\bibfnamefont {O.}~\bibnamefont {{Romero-Isart}}},\ and\ \bibinfo {author} {\bibfnamefont {L.}~\bibnamefont {Novotny}},\ }\bibfield  {title} {\bibinfo {title} {Cavity-mediated long-range interactions in levitated optomechanics},\ }\href {https://doi.org/10.1038/s41567-024-02405-3} {\bibfield  {journal} {\bibinfo  {journal} {Nature Physics}\ }\textbf {\bibinfo {volume} {20}},\ \bibinfo {pages} {859} (\bibinfo {year} {2024})}\BibitemShut {NoStop}%
\bibitem [{\citenamefont {Xu}\ \emph {et~al.}(2023)\citenamefont {Xu}, \citenamefont {Zheng}, \citenamefont {Wang}, \citenamefont {Zoller}, \citenamefont {Clerk},\ and\ \citenamefont {Jiang}}]{xuAutonomousQuantumError2023}%
  \BibitemOpen
  \bibfield  {author} {\bibinfo {author} {\bibfnamefont {Q.}~\bibnamefont {Xu}}, \bibinfo {author} {\bibfnamefont {G.}~\bibnamefont {Zheng}}, \bibinfo {author} {\bibfnamefont {Y.-X.}\ \bibnamefont {Wang}}, \bibinfo {author} {\bibfnamefont {P.}~\bibnamefont {Zoller}}, \bibinfo {author} {\bibfnamefont {A.~A.}\ \bibnamefont {Clerk}},\ and\ \bibinfo {author} {\bibfnamefont {L.}~\bibnamefont {Jiang}},\ }\bibfield  {title} {\bibinfo {title} {Autonomous quantum error correction and fault-tolerant quantum computation with squeezed cat qubits},\ }\href {https://doi.org/10.1038/s41534-023-00746-0} {\bibfield  {journal} {\bibinfo  {journal} {npj Quantum Information}\ }\textbf {\bibinfo {volume} {9}},\ \bibinfo {pages} {78} (\bibinfo {year} {2023})}\BibitemShut {NoStop}%
\bibitem [{\citenamefont {Fruchart}\ \emph {et~al.}(2021)\citenamefont {Fruchart}, \citenamefont {Hanai}, \citenamefont {Littlewood},\ and\ \citenamefont {Vitelli}}]{fruchartNonreciprocalPhaseTransitions2021}%
  \BibitemOpen
  \bibfield  {author} {\bibinfo {author} {\bibfnamefont {M.}~\bibnamefont {Fruchart}}, \bibinfo {author} {\bibfnamefont {R.}~\bibnamefont {Hanai}}, \bibinfo {author} {\bibfnamefont {P.~B.}\ \bibnamefont {Littlewood}},\ and\ \bibinfo {author} {\bibfnamefont {V.}~\bibnamefont {Vitelli}},\ }\bibfield  {title} {\bibinfo {title} {Non-reciprocal phase transitions},\ }\href {https://doi.org/10.1038/s41586-021-03375-9} {\bibfield  {journal} {\bibinfo  {journal} {Nature}\ }\textbf {\bibinfo {volume} {592}},\ \bibinfo {pages} {363} (\bibinfo {year} {2021})}\BibitemShut {NoStop}%
\bibitem [{\citenamefont {Zhu}\ \emph {et~al.}(2024)\citenamefont {Zhu}, \citenamefont {Hu}, \citenamefont {Wang}, \citenamefont {Qin}, \citenamefont {L{\"u}},\ and\ \citenamefont {Nori}}]{zhuNonreciprocalSuperradiantPhase2024}%
  \BibitemOpen
  \bibfield  {author} {\bibinfo {author} {\bibfnamefont {G.-L.}\ \bibnamefont {Zhu}}, \bibinfo {author} {\bibfnamefont {C.-S.}\ \bibnamefont {Hu}}, \bibinfo {author} {\bibfnamefont {H.}~\bibnamefont {Wang}}, \bibinfo {author} {\bibfnamefont {W.}~\bibnamefont {Qin}}, \bibinfo {author} {\bibfnamefont {X.-Y.}\ \bibnamefont {L{\"u}}},\ and\ \bibinfo {author} {\bibfnamefont {F.}~\bibnamefont {Nori}},\ }\bibfield  {title} {\bibinfo {title} {Nonreciprocal {{Superradiant Phase Transitions}} and {{Multicriticality}} in a {{Cavity QED System}}},\ }\href {https://doi.org/10.1103/PhysRevLett.132.193602} {\bibfield  {journal} {\bibinfo  {journal} {Physical Review Letters}\ }\textbf {\bibinfo {volume} {132}},\ \bibinfo {pages} {193602} (\bibinfo {year} {2024})}\BibitemShut {NoStop}%
\bibitem [{\citenamefont {Zhang}\ \emph {et~al.}(2025)\citenamefont {Zhang}, \citenamefont {Lin}, \citenamefont {Feng}, \citenamefont {Kang},\ and\ \citenamefont {Xiong}}]{zhangNonreciprocalSuperradiantQuantum2025}%
  \BibitemOpen
  \bibfield  {author} {\bibinfo {author} {\bibfnamefont {G.-Q.}\ \bibnamefont {Zhang}}, \bibinfo {author} {\bibfnamefont {S.-Y.}\ \bibnamefont {Lin}}, \bibinfo {author} {\bibfnamefont {W.}~\bibnamefont {Feng}}, \bibinfo {author} {\bibfnamefont {Y.-H.}\ \bibnamefont {Kang}},\ and\ \bibinfo {author} {\bibfnamefont {W.}~\bibnamefont {Xiong}},\ }\href {https://doi.org/10.48550/arXiv.2509.25985} {\bibinfo {title} {Nonreciprocal superradiant quantum phase transition induced by magnon {{Kerr}} effect}} (\bibinfo {year} {2025}),\ \Eprint {https://arxiv.org/abs/2509.25985} {arXiv:2509.25985 [quant-ph]} \BibitemShut {NoStop}%
\end{thebibliography}%
\clearpage

\onecolumngrid
\pagebreak


\section*{Supplementary material}

# Supplementary Material for
# Higher symmetry breaking and non-reciprocity in a driven-dissipative Dicke model

Jacquelyn Ho,[1,2] Yue-Hui Lu,[1,2] Tai Xiang,[1,2] Tsai-Chen Lee,[1,2] Zhenjie Yan,[1,2] and Dan M. Stamper-Kurn[1,2,3,*]

[1]*Department of Physics, University of California, Berkeley, California 94720*
[2]*Challenge Institute for Quantum Computation, University of California, Berkeley, California 94720*
[3]*Materials Sciences Division, Lawrence Berkeley National Laboratory, Berkeley, California 94720*

## I. STEADY-STATE STABILITY ANALYSIS

### A. Lyapunov function for the $\kappa = 0$ case

Lyapunov functions provide a convenient way for finding stable equilibria of dynamical systems. If a Lyapunov function $L$ exists (i.e. it satisfies $\partial L/\partial p_j = \dot{z}_j$ and $\partial L/\partial z_j = -\dot{p}_j$), equilibria correspond to local minima of $L$, which are Lyapunov stable [S1]. In the $\kappa = 0$ case, we construct a Lyapunov function that is given by

$$L = \sum_{j=1}^{n} \left( \frac{p_j^2}{2\mu} + \frac{\mu \omega_z^2 z_j^2}{2} + C \Big[ -\frac{\Delta_{\mathrm{pc}} \cos(2kz_j)}{4k} + \sum_{l<j} \frac{\sin(kz_j)}{k} \sin(kz_l)\big(\Delta_{\mathrm{pc}} \cos(\phi(j-l))\big) \Big] \right), \tag{S1}$$

where we have defined $C = \frac{2\hbar\nu^2\Omega^2}{\Delta_{\mathrm{pa}}^2} \frac{g_0^2 k}{\Delta_{\mathrm{pc}}^2 + \kappa^2}$. We note that for $n = 1$ and $n = 2$, $L$ is also an exact Lyapunov function even if $\kappa > 0$, since the term proportional to $\kappa$ in the sum in Eq. (6) of the main text vanishes when $\phi = 2\pi$ and $\phi = \pi$. To obtain the plots in Fig. 1(c), we minimize $L$ for $n = 1, 2, ..., 6$ using simplicial homology global optimization [S2] with the following parameters: $\nu = 30$, $\omega_z = 2\pi \times 70$ kHz, $\Delta_{\mathrm{pa}} = -2\pi \times 100$ MHz, $\Delta_{\mathrm{pc}} = -2\pi \times 4$ MHz, $g_0 = 2\pi \times 3$ MHz, and $\Omega = 2\pi \times 20$ MHz. Plugging the solutions $z_j$ into Eq. (5) reveals a $\mathbb{Z}_n$ symmetry for even $n$ and a $\mathbb{Z}_{2n}$ symmetry for odd $n$.

### B. Jacobian eigenvalues

To evaluate the linear stability of the steady states shown in Fig. 2, we compute the Jacobian matrix $\mathbf{J}$ and its eigenvalues evaluated at each steady state. The elements of $\mathbf{J}$ are given by $J_{ij} = \partial F_i/\partial x_j$, where $\mathbf{x} = (\mathrm{Re}[c], \mathrm{Im}[c], p_1, ..., p_n, z_1, ..., z_n)^T$ and $\mathbf{F}(\mathbf{x})$ is given by the right-hand-side of Eqs. (2-4). Any eigenvalue $\lambda_j$ of $\mathbf{J}$ with a positive real part indicates instability.

In Fig. S1, we show the absolute value of the steady-state positions for one group $|z_j|$ at a particular symmetry-broken solution and the fraction of the Jacobian eigenvalues $\lambda_j$ with positive or negative real part evaluated at that steady state for each point in the parameter space spanned by $\Omega$ and $\Delta_{\mathrm{pc}}$ for $n = 3 - 6$. For all $n$, we see that there is a parameter regime where no eigenvalues have a positive real part. For odd $n$, all eigenvalues have negative real part in that region, indicating asymptotic stability, whereas for even $n$ some of the eigenvalues are purely imaginary. This indicates that there are some eigenmodes of the even $n$ systems that are not subject to cavity cooling. From this linear stability analysis, one cannot conclude that the even $n$ steady states are stable because of the purely imaginary eigenvalues. However, from trajectories it appears that the symmetry-broken states are indeed stable.

## II. APPROXIMATION OF PHASE BOUNDARIES

To obtain an analytical approximation for the phase boundaries, we expand Eq. (6) of the main text to third order around a symmetry-broken solution where the atoms have self-organized onto the antinodes of the cavity field.

---

* dmsk@berkeley.edu

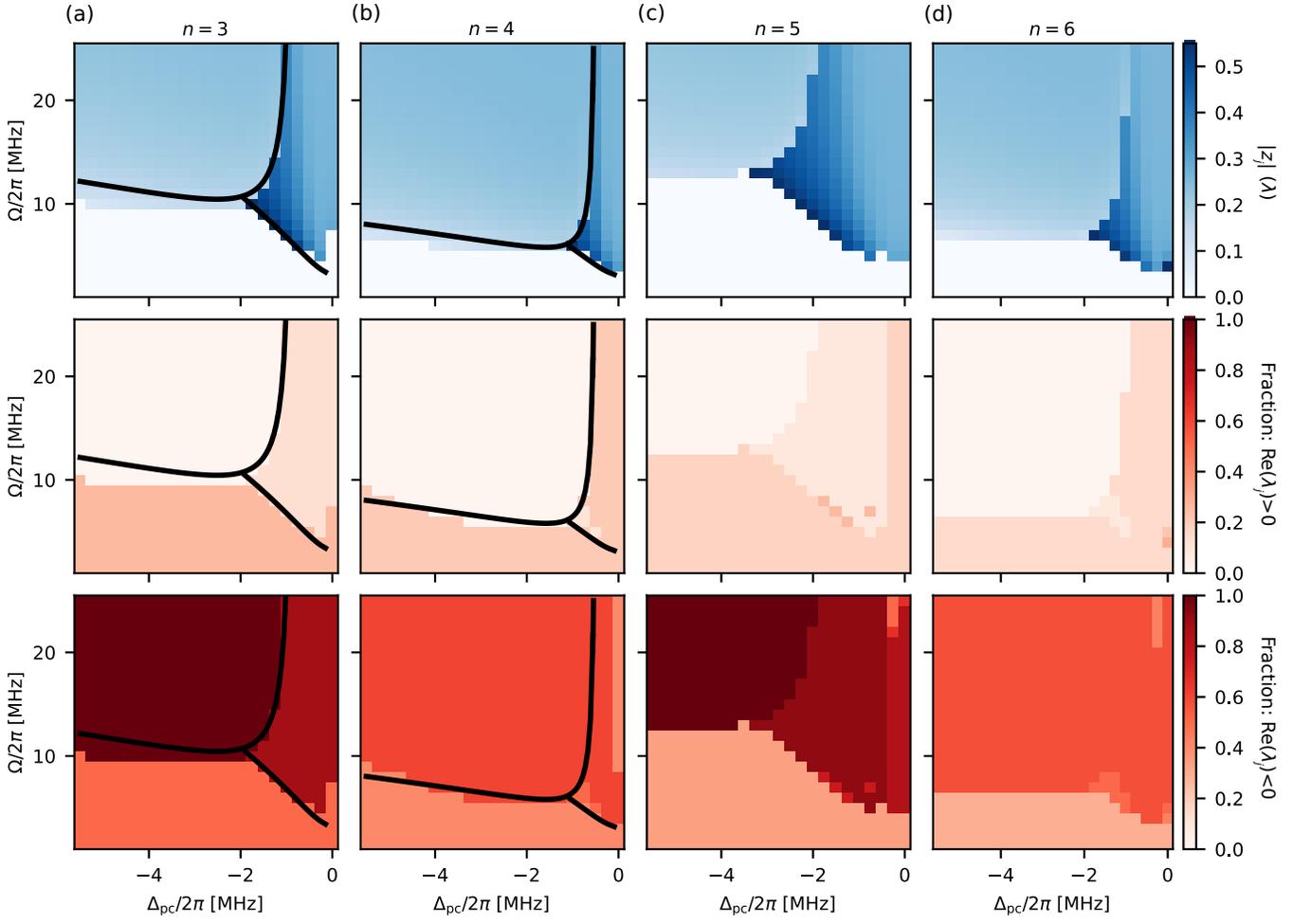

Figure S1. (a-d) Phase diagrams of the $n$-phase Dicke model for $n = 3 - 6$. Top row (blue) shows the absolute value of the steady-state displacement of the stability-determining group (e.g. $|z_2|$ in Eqs. (S5-S8) for the $n = 3$ case) or pair of groups (for even $n$). The middle and bottom rows show the fraction of Jacobian eigenvalues with positive or negative real part. We see that for $n = 3$ and $n = 5$, the stable symmetry-broken region is characterized by eigenvalues that all have a negative real part, indicating asymptotic stability. On the other hand, $n = 4$ and $n = 6$ each have $n$ eigenvalues that are purely imaginary in the stable symmetry-broken region. Stable symmetry breaking is thus confirmed by examining time trajectories. The phase boundaries drawn in (a) and (b) are based on the analysis described in Section II.

First, let us consider $n = 3$. The force equations are

$$\dot{p}_1 = -\mu\omega_z^2 z_1 - C\cos(kz_1)\left[\Delta_{\text{pc}}\sin(kz_1) + \left(-\frac{\Delta_{\text{pc}}}{2} + \frac{\sqrt{3}\kappa}{2}\right)\sin(kz_2) + \left(-\frac{\Delta_{\text{pc}}}{2} - \frac{\sqrt{3}\kappa}{2}\right)\sin(kz_3)\right], \quad \text{(S2)}$$

$$\dot{p}_2 = -\mu\omega_z^2 z_2 - C\cos(kz_2)\left[\Delta_{\text{pc}}\sin(kz_2) + \left(-\frac{\Delta_{\text{pc}}}{2} + \frac{\sqrt{3}\kappa}{2}\right)\sin(kz_3) + \left(-\frac{\Delta_{\text{pc}}}{2} - \frac{\sqrt{3}\kappa}{2}\right)\sin(kz_1)\right], \quad \text{(S3)}$$

$$\dot{p}_3 = -\mu\omega_z^2 z_3 - C\cos(kz_3)\left[\Delta_{\text{pc}}\sin(kz_3) + \left(-\frac{\Delta_{\text{pc}}}{2} + \frac{\sqrt{3}\kappa}{2}\right)\sin(kz_1) + \left(-\frac{\Delta_{\text{pc}}}{2} - \frac{\sqrt{3}\kappa}{2}\right)\sin(kz_2)\right]. \quad \text{(S4)}$$

We now expand Eqs. (S2-S4) around the symmetry-broken state $z_{1,0} = +\lambda/4$, $z_{2,0} = -\lambda/4$, $z_{3,0} = -\lambda/4$ and examine

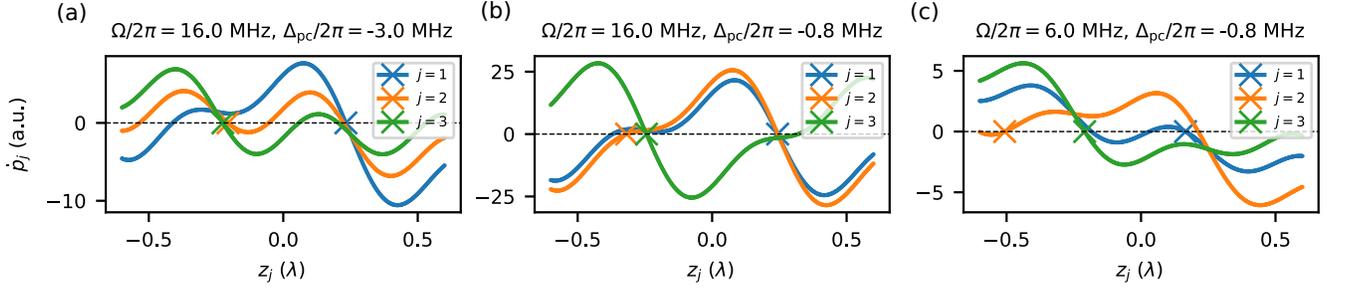

Figure S2. The force $\dot{p}_j$ on each group as a function of its displacement $z_j$ assuming that the other groups are pinned at a particular symetry-broken steady-state solution (marked by "x"'s), shown at three different parameter choices for $n = 3$. In (a), the system is in the stable symmetry breaking regime, and we see that the slope of the force is negative around each steady state and is thus a restoring force. In (b) and (c), the system is in the unstable broken symmetry phase, and we see that the slope of $\dot{p}_2(z_2)$ is positive around the $z_2$ steady state, indicating instability. The phase boundaries occur at parameters where the local minima of $\dot{p}_2$ (such as the one to the right of the orange x in (a) and the one to the left of the orange x in (c)) touch the $\dot{p}_2 = 0$ line.

the force for displacements $\delta z_j$ away from that state. Doing so yields

$$\dot{p}_1 = -\mu\omega_z^2\left(\frac{\lambda}{4} + \delta z_1\right) + C\left(2\Delta_{\text{pc}}k\delta z_1 - \frac{5}{6}\Delta_{\text{pc}}k^3\delta z_1^3\right) + \mathcal{O}(\delta z_1^5) + \mathcal{O}(\delta z_2^2) + \mathcal{O}(\delta z_3^2), \tag{S5}$$

$$\dot{p}_2 = -\mu\omega_z^2\left(-\frac{\lambda}{4} + \delta z_2\right) + C\left((\sqrt{3}\kappa + \Delta_{\text{pc}})k\delta z_2 - \left(\frac{2}{3}\Delta_{\text{pc}} + \frac{\sqrt{3}}{6}\kappa\right)k^3\delta z_2^3\right) + \mathcal{O}(\delta z_2^5) + \mathcal{O}(\delta z_1^2) + \mathcal{O}(\delta z_3^2), \tag{S6}$$

$$\dot{p}_3 = -\mu\omega_z^2\left(-\frac{\lambda}{4} + \delta z_3\right) + C\left((\Delta_{\text{pc}} - \sqrt{3}\kappa)k\delta z_3 - \left(\frac{2}{3}\Delta_{\text{pc}} - \frac{\sqrt{3}}{6}\kappa\right)k^3\delta z_2^3\right) + \mathcal{O}(\delta z_3^5) + \mathcal{O}(\delta z_1^2) + \mathcal{O}(\delta z_2^2). \tag{S7}$$

Since $\Delta_{\text{pc}} < 0$ and $\kappa > 0$, it becomes apparent that in Eq. (S6), the terms proportional to $\kappa$ negate the effect of the terms proportional to $\Delta_{\text{pc}}$. This is not the case in Eq. (S5), where the $\kappa$ terms drop out, or Eq. (S7), where the $\kappa$ and $\Delta_{\text{pc}}$ terms have the same sign. Therefore, we expect that Eq. (S6) determines the stability of the system due to the presence of the reactive $\kappa$ terms. Examining the full equations for the force in Fig. S2, this appears to be the case. Fig. S2(a) shows the force vs. $z_j$ in the dispersive, stable symmetry-broken regime, where each "x" marks the steady-state solution for each $j$. The force around each steady state has a negative slope, indicating that the force is restoring and that the system is stable. However, in Fig. S2(b-c) we are in the reactive regime; we see that $\dot{p}_2$ no longer has a local minimum near $z_2 = -\lambda/4$ and now has a positive slope around the steady state for $z_2$. This shows that $z_2$ is in an unstable steady state.

Fig. S2(b) also shows that the steady state of $z_2$ moves beyond the antinode at $-\lambda/4$. When lowering the pump strength, this steady state is pushed even farther and becomes near $-\lambda/2$, while $z_1$ and $z_3$ remain closer to the antinodes [Fig. S2(c)]. One can expand the equation for $\dot{p}_2$ around $z_{2,0} = -\lambda/2$ and obtain

$$\dot{p}_2 = -\mu\omega_z^2\left(-\frac{\lambda}{2} + \delta z_2\right) + C\left(-\sqrt{3}\kappa - \Delta_{\text{pc}}k\delta z_2 + \frac{\sqrt{3}}{2}\kappa k^2\delta z_2^2 + \frac{2}{3}\Delta_{\text{pc}}k^3\delta z_2^3\right) + \mathcal{O}(\delta z_2^4) + \mathcal{O}(\delta z_1^2) + \mathcal{O}(\delta z_3^2). \tag{S8}$$

Dropping the higher order terms in Eqs. (S6) and (S8), one can obtain an approximate solution for the phase boundaries. The condition for the boundaries is that $\dot{p}_2(z_{2,\text{min}}) = 0$ where $z_{2,\text{min}}$ is a local minimum of $\dot{p}_2$. If $\dot{p}_2(z_{2,\text{min}}) > 0$, then there is no steady state in the vicinity of $z_{2,\text{min}}$. In the vicinity of $z_{2,0} = -\lambda/4$, the local minimum of Eq. (S6) is given by $z_{2,\text{min}} = -\lambda/4 + \sqrt{2}\sqrt{(Ck\Delta_{\text{pc}} + \sqrt{3}Ck\kappa - \mu\omega_z^2)/(4Ck^3\Delta_{\text{pc}} + \sqrt{3}Ck^3\kappa)}$ and in the vicinity of $z_{2,0} = -\lambda/2$, the local minimum of Eq. (S8) is $z_{2,\text{min}} = -\lambda/2 + (-\sqrt{3}Ck^2\kappa + \sqrt{C}k^{3/2}\sqrt{8Ck\Delta_{\text{pc}}^2 + 3Ck\kappa^2 + 8\Delta_{\text{pc}}\mu\omega^2})/(4Ck^3\Delta_{\text{pc}})$. Numerically solving $z_{2,\text{min}} = 0$ in each of these cases, in the appropriate regions of the $\Omega$ and $\Delta_{\text{pc}}$ parameter space, gives the boundaries shown in Fig. 2(b) and Fig. S1(a).

A similar analysis was performed for $n = 4$ to obtain the phase boundaries shown in Fig. S1(b). When $n$ is even, the number of force equations can be reduced by half using the symmetry $z_j = -z_{j+n/2}$. We expect this analysis also to extend to $n > 4$. Discrepancies between the drawn phase boundaries and the results of the numerically solved steady states likely originate from the higher order terms in Eqs. (S5-S8) that were dropped.

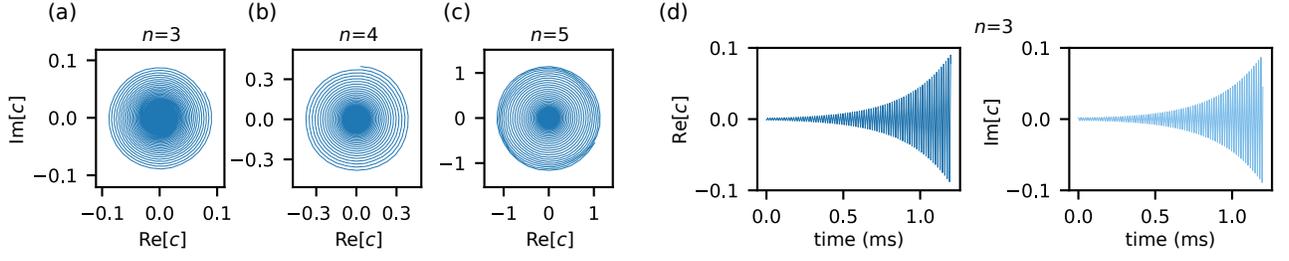

Figure S3. Cavity field trajectories in the weak pump limit for (a) $n = 3$, (b) $n = 4$, and (c) $n = 5$. Normal mode analysis predicts that the $c = 0$ state is unstable for $\kappa > 0$ and $\Omega > 0$; this is consistent with the trajectories, which spiral outward from $c = 0$. All trajectories are integrated over a time span of 1.2 ms. (d) Time trajectories of the real and imaginary quadratures of $c$ for $n = 3$, corresponding to the plot shown in (a). Parameters used are $\Omega = 2\pi \times 2$ MHz, $\Delta_{\rm pc} = -2\pi \times 4$ MHz, $\Delta_{\rm pa} = -2\pi \times 100$ MHz, $\kappa = 2\pi \times 0.5$ MHz, $\nu = 30$, $g_0 = 2\pi \times 3$ MHz, and $\omega_z = 2\pi \times 70$ kHz.

## III. INSTABILITY OF THE UNBROKEN-SYMMETRY STATE

We seek to understand the stability of the unbroken-symmetry solution. For this, we make the Lamb-Dicke approximation $k|z_j| \ll 1$ and examine the mechanical modes. Under this approximation, Eq. (6) of the main text can be reduced to a linear equation

$$\mu \ddot{z}_j = -\mu \omega_z^2 z_j - Ck \sum_{l=1}^{n} \left[ \Delta_{\rm pc} \cos(\phi(j - l)) - \kappa \sin(\phi(j - l)) \right] z_l \tag{S9}$$

that can be solved by seeking a solution of the form $\tilde{\mathbf{z}}(t) = \tilde{\mathbf{z}} e^{-i\omega t}$, where $\tilde{\mathbf{z}} = (z_1, z_2, ..., z_n)^T$. This leads to the equation

$$\omega^2 \tilde{\mathbf{z}} = \omega_z^2 \tilde{\mathbf{z}} + \frac{2\hbar k^2 g_0^2 \nu^2 \Omega^2}{\mu \Delta_{\rm pa}^2 \sqrt{\Delta_{\rm pc}^2 + \kappa^2}} \left( \cos\theta \mathbf{M}_{\rm c} + \sin\theta \mathbf{M}_{\rm s} \right) \tilde{\mathbf{z}}, \tag{S10}$$

where we have defined $\theta$ such that $\tan\theta = -\kappa/\Delta_{\rm pc}$ (i.e. $\theta$ is the phase response of the cavity), and we introduce the matrices $\mathbf{M}_{\rm c}$ and $\mathbf{M}_{\rm s}$ that have matrix elements $M_{{\rm c},jl} = \cos(\phi(j-l))$ and $M_{{\rm s},jl} = \sin(\phi(j-l))$. $\mathbf{M}_{\rm c}$ and $\mathbf{M}_{\rm s}$ both exhibit $\mathbb{Z}_n$ symmetry and have $n$ eigenvectors $\tilde{\mathbf{z}}^{(q)}$ whose elements are $\tilde{z}_j^{(q)} = e^{i\phi jq}/\sqrt{n}$, where $q = 1, 2, ..., n$. It can be shown that when $n > 2$, $\mathbf{M}_{\rm c} \tilde{\mathbf{z}}^{(q)} = \mathbf{M}_{\rm s} \tilde{\mathbf{z}}^{(q)} = 0$ for all $q$ except $q = 1$ and $q = n - 1$; we label these two eigenvectors as $\tilde{\mathbf{z}}^{(+)}$ and $\tilde{\mathbf{z}}^{(-)}$ respectively. We find that for the $n - 2$ modes with eigenvalue 0, the eigenfrequencies are simply given by $\omega^2 = \omega_z^2$, whereas for the $\tilde{\mathbf{z}}^{(\pm)}$ modes we have

$$\omega^2 \tilde{\mathbf{z}}^{(\pm)} = \left[ \omega_z^2 + \frac{N\Omega^2 \hbar k^2 g_0^2}{m \Delta_{\rm pa}^2 \sqrt{\Delta_{\rm pc}^2 + \kappa^2}} \left( \cos\theta \pm i \sin\theta \right) \right] \tilde{\mathbf{z}}^{(\pm)}. \tag{S11}$$

We immediately see that for $\kappa > 0$, the eigenfrequencies have a nonzero imaginary part, indicating exponentially growing and exponentially damped eigenmodes. The presence of exponentially growing modes leads the $\mathbf{z} = 0$ solution to be unstable for any $\Omega > 0$.

## IV. TRAJECTORIES IN THE WEAK PUMP LIMIT

The instability of the unbroken symmetry phase caused by non-reciprocal interactions is evident in the dynamics of the cavity field for $n = 3, 4$, and $5$, obtained by integrating Eqs. (2-4) of the main text and shown in Fig. S3. Given a small perturbation of the atoms' initial positions away from $z_j = 0$, the cavity field spirals away from the $c = 0$ state, consistent with the dynamics of an exponentially growing mode. As the atom positions get farther away from the cavity field nodes, their dynamics becomes dictated by the form of the nonlinearity in the system. In this case, where there is a sinusoidal cavity field profile, simulating trajectories out to later times reveals limit-cycle-like behavior where the atoms oscillate between the cavity nodes located at $\pm \lambda/2$.

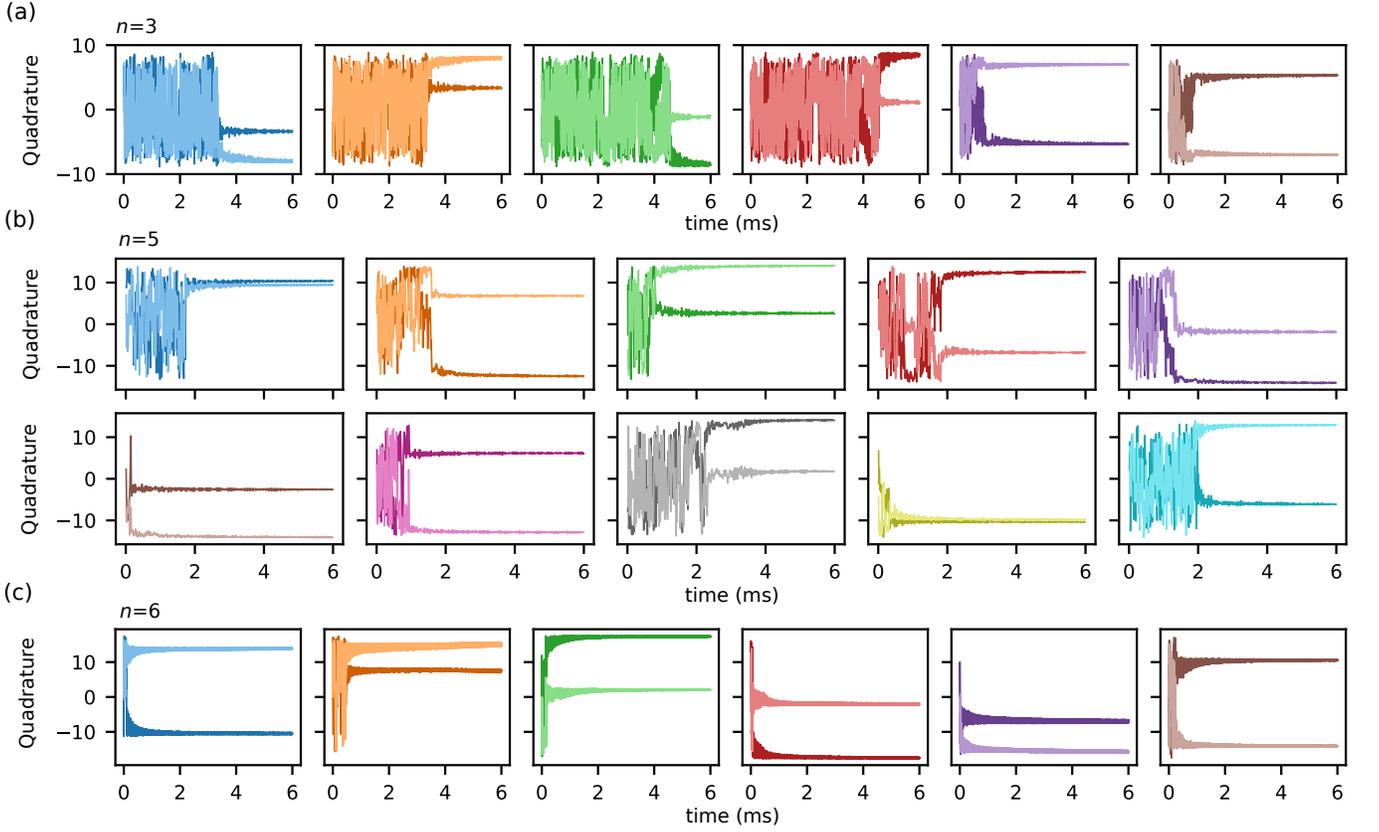

Figure S4. $n = 3$ (a), $n = 5$ (b), and $n = 6$ (c) cavity field timetraces for the parameters given in Fig. 2 of the main text. Different colors correspond to different initial perturbations around the $z_j = 0$, $p_j = 0$ steady state. Dark colors indicate the real part of the cavity field and light colors indicate the imaginary part of the cavity field.

## V. BROKEN SYMMETRY CAVITY FIELD TRAJECTORIES

As shown in Fig. 3 of the main text, the cavity field trajectories display symmetry breaking at long times, at which the trajectories converge at the vertices of a polygon. This is shown explicitly in Fig. 3(d) for $n = 4$. In Fig. S4 we show similar timetraces of the real and imaginary quadratures of the cavity field for $n = 3$, $n = 5$, and $n = 6$ for different initial perturbations around the $z_j = 0$, $p_j = 0$ steady state. The odd $n$ trajectories appear to take longer time to reach the symmetry broken states than in the even $n$ case.

## VI. EFFECT OF ATOMIC SATURATION

The model introduced in Eq. (1) of the main text assumes adiabatic elimination of the atomic excited state. When including the excited state dynamics (including its decay with half-linewidth $\gamma$), the system can still break symmetry stably in some parameter regime [Fig. S5(a)]. However, it appears that there is also a regime where the excited state dynamics makes the center-of-mass motion unstable when the cavity field is too large or the atomic excited state does not dissipate rapidly enough [Figs. S5(b)]. This instability can be mitigated in a system of fewer atoms, with the caveat that the atom number must still be high enough to reach the self-organized phase.

## VII. EFFECT OF TEMPERATURE

Temperature affects the onset of optomechanical self-organization because fluctuations in the atom positions and momenta cause the excitation of mechanical normal modes within each sub-ensemble that do not couple to the cavity. The dominant mode (the mode that is linearly coupled to the cavity, which is the center-of-mass mode in our model)

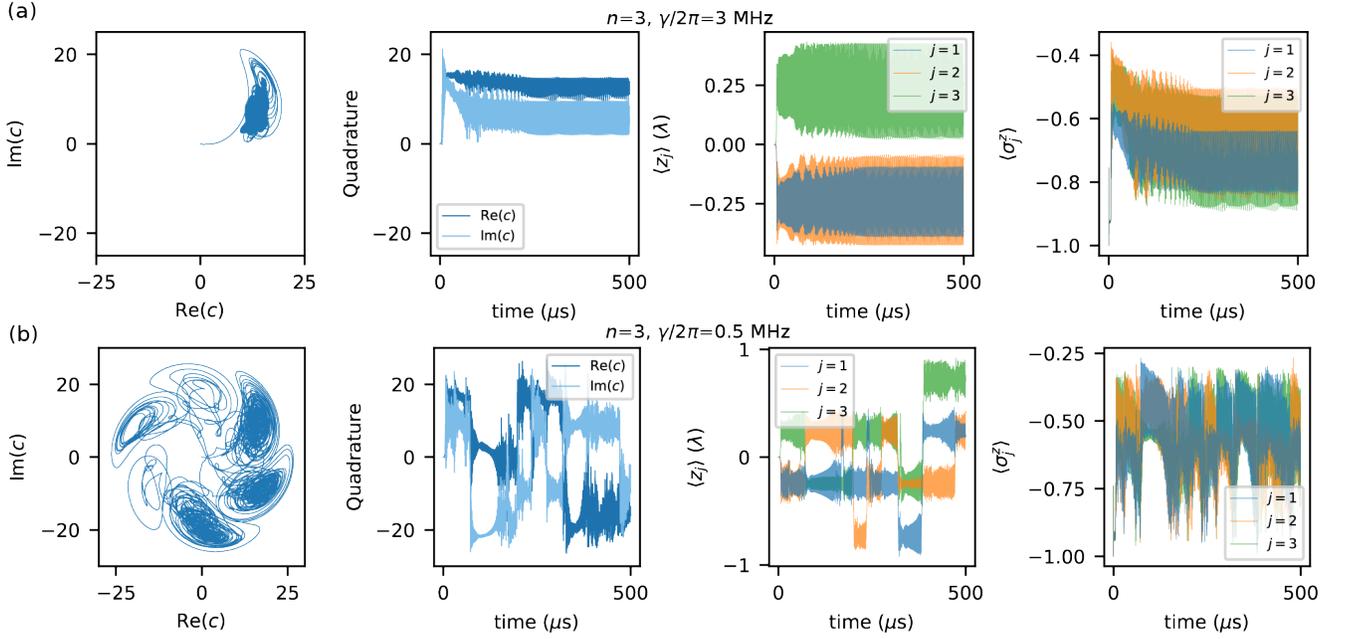

Figure S5. Trajectories of the cavity field, atomic positions, and atomic spins for $n = 3$ with (a) $\gamma = 2\pi \times 3$ MHz and (b) $\gamma = 2\pi \times 0.5$ MHz. Other parameters are the same as in Fig. 2 of the main text. Note that here, only the center-of-mass mode of each group is excited. In (a) the system breaks symmetry stably and on a timescale of $\sim 100$ $\mu$s, while in (b) the system does not appear to break symmetry.

is nonlinearly coupled to these non-dominant modes, which effectively serve as a thermal bath for the dominant mode [S3].

To simulate the effect of temperature $T$, we sample the positions and momenta of all $N$ atoms from a thermal Boltzmann distribution defined by $T$. When doing so, it appears that at large atom numbers, instabilities arise from the nonlinear coupling of the dominant mode to the other modes. As shown in Fig. S6(a), the system does not break symmetry stably at a temperature of $T = 3$ $\mu$K for $n = 3$ and $\nu = 30$. Rather, the atoms seem to approach a symmetry-broken solution at early times but then evolve into chaotic motion centered around the cavity nodes. However, the system does appear to break symmetry when going to a lower atom number ($\nu = 15$) even at $T = 10$ $\mu$K [Fig. S6(b-c)]. It is surprising that stability occurs at lower atom number; investigating this will be the subject of future work. We speculate that the reasons for this behavior could be related to why we see instability at lower $\gamma$, as shown in Fig. S5(b).

While in the atomic system, it seems that temperature complicates symmetry-breaking dynamics, it is possible that this issue can be circumvented in other platforms where non-dominant modes can be eliminated. For instance, if one could replace each group of atoms with a single mechanical element that can be cooled to its motional ground state, it may be possible to restore symmetry-breaking dynamics. Nonetheless, our simulations indicate that selecting specific system parameters (including atom number) allows for symmetry-broken solutions to be realized.

---

[S1] D. G. Schaeffer and J. W. Cain, *Ordinary Differential Equations: Basics and Beyond*, Texts in Applied Mathematics (Springer New York, New York, NY, 2016).
[S2] S. C. Endres, C. Sandrock, and W. W. Focke, A simplicial homology algorithm for Lipschitz optimisation, Journal of Global Optimization **72**, 181 (2018).
[S3] J. Ho, Y.-H. Lu, T. Xiang, C. C. Rusconi, S. J. Masson, A. Asenjo-Garcia, Z. Yan, and D. M. Stamper-Kurn, Optomechanical self-organization in a mesoscopic atom array, Nature Physics **21**, 1071 (2025).

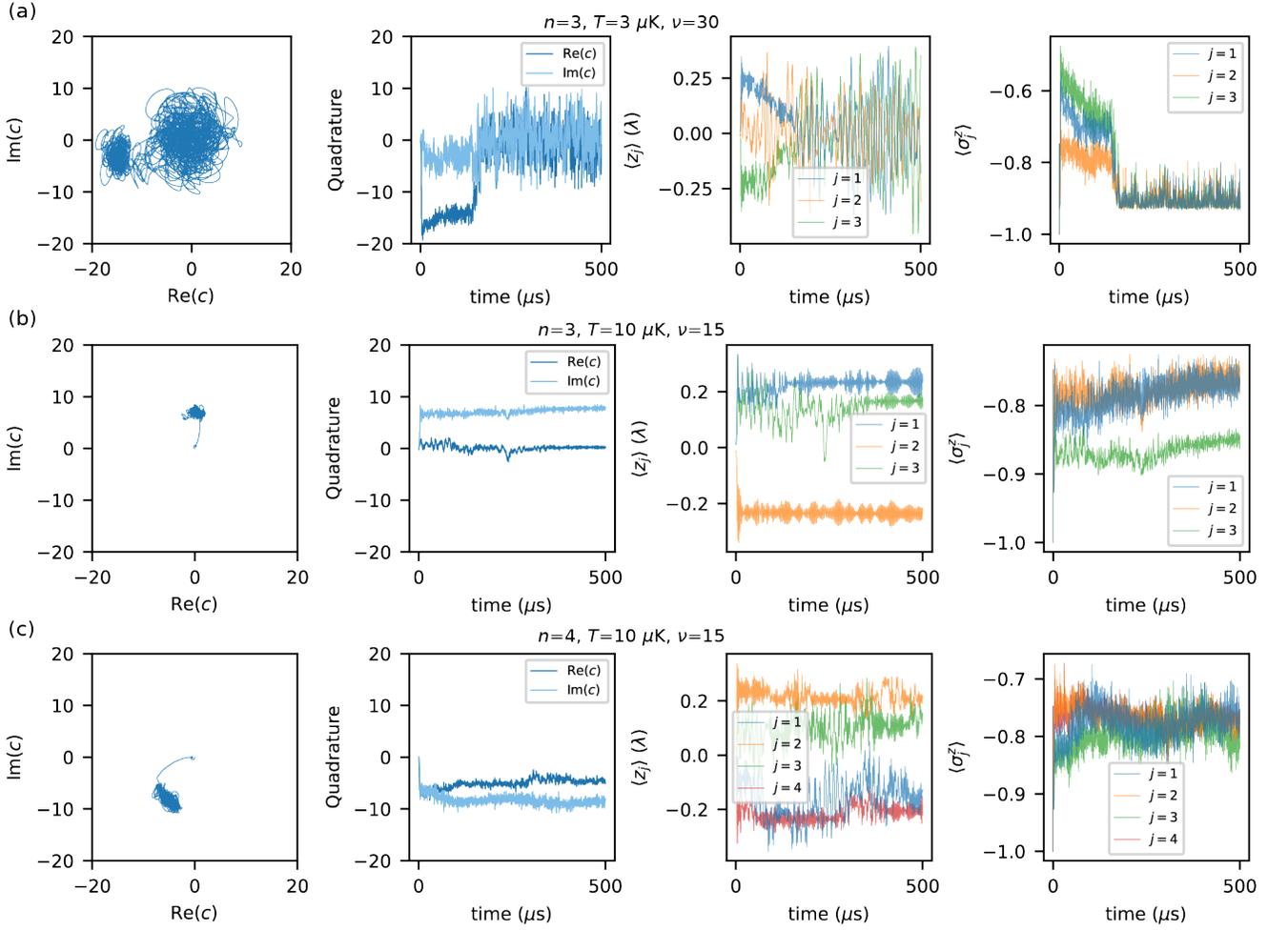

Figure S6. (a) Using the same parameters as in Fig. S5(a) but giving the atoms a temperature of 3 $\mu$K, we see that the system no longer breaks symmetry. However, by reducing the number of atoms in each group to $\nu = 15$ but otherwise using the same parameters as in Fig. S5(a), we see that symmetry breaking is restored in the system for both $n = 3$ (b) and $n = 4$ (c) even at a temperature of $T = 10$ $\mu$K.


\end{document}